\documentclass[APA,STIX1COL]{WileyNJD-v2}

\newcommand{\bbeta}{\mbox{\boldmath $\beta$}}

\newcommand{\bTheta}{\mbox{\boldmath $\Theta$}}
\newcommand{\bxi}{\mbox{\boldmath $\xi$}}

\newcommand{\bDelta}{\mbox{\boldmath $\Delta$}}

\newcommand{\bvartheta}{\mbox{\boldmath $\vartheta$}}

\usepackage{float} 
\usepackage{lscape} 
\usepackage{booktabs} 
\usepackage{placeins}
\usepackage{graphicx}

\articletype{RESEARCH ARTICLE}%

\received{xx xx 2024}
\revised{xx xx 2024}
\accepted{xx xx 2024}

\begin{document}

\title{A Bayesian Spatial-Temporal Functional Model for Data with Block Structure and Repeated Measures}

\author[1]{David H. da Matta}

\author[2]{Mariana R. Motta}

\author[3]{Nancy L. Garcia}

\author[4]{Alexandre B. Heinemann}

\authormark{AUTHOR ONE \textsc{et al}}

\address[1]{\orgdiv{Institute of Mathematics and Statistics}, \orgname{Universidade3 Federal de Goiás (UFG)}, \orgaddress{\state{Goiás}, \country{Brazil}}}

\address[2]{\orgdiv{Institute of Mathematics, Statistics, and Scientific Computing}, \orgname{Universidade Estadual de Campinas (UNICAMP)}, \orgaddress{\state{São Paulo}, \country{Brazil}}}

\address[3]{\orgdiv{Institute of Mathematics, Statistics, and Scientific Computing}, \orgname{Universidade Estadual de Campinas (UNICAMP)}, \orgaddress{\state{São Paulo}, \country{Brazil}}}

\address[4]{\orgname{Embrapa Arroz e Feijão}, \orgaddress{\state{Goiás}, \country{Brazil}}}

\corres{*Mariana R. Motta,  \email{marirm@unicamp.br}}


\abstract[Summary]{ The analysis of spatio-temporal data has been the object of research in several areas of knowledge. One of the main objectives of such research is the need to evaluate the behavior of climate effects in certain regions across a period of time. When certain climate patterns appear for several days or even weeks, causing the areas affected by them to have the same kind of weather for an extended period of time, the use of blocks for these phenomena may be a good strategy. Additionally, having repeated measures for observations within blocks helps to control for differences between observations, thus gaining more statistical power. In view of these perspectives, this study presents a spatio-temporal regression model with block structure with repeated measures incorporating as predictors functional variables of fixed and random nature. To accommodate complex spatial, temporal and block structures, functional components based on random effects were considered in addition to the class {\itshape{Matérn}} covariance structure,  which was responsible to account for spatial covariance. This work is motivated by a precipitation dataset collected monthly from various meteorological stations in Goiás State, Brazil, covering the years 1980 to 2001 (21 years).  In this framework, spatial effects are represented by individual meteorological stations, temporal effects by months, block effects by climate patterns, and repeated measures by the years within those patterns. The proposed model demonstrated promising results in simulation studies and effectively estimated precipitation using the available data.}

\keywords{Functional data analysis, Multivariate spatio-temporal models, Bayesian inference, Markov Chain Monte Carlo.}


\maketitle


\section{Introduction}\label{sect:intro}  

Rainfed cropping is the main grain production system in Brazil. In this system, all soil moisture is supplied by precipitation. As plants extract water from the soil, any change in the amount of precipitation will affect the supply of soil moisture, thereby affecting the final crop yield. With the advent of ``enviromics'', defined as a process dealing with environmental characterization by envirotyping for micro or macro-environments \citep{resende2022enviromics}, the demand for available and reliable meterological data, mainly precipitation, became increasingly important. Many studies have shown the importance of linking agronomic data (yield, flowering date, harvest date, etc.) with meterological data for a better understanding of the interaction of crop growth and environment \citep{heinemann2024climate,costa2023environmental,heinemann2022enviromic}. However, Brazil is a continental country and there are large areas without any rain gauges or weather stations, the main sources of weather data \citep{menezes2022impact,xavier2016daily}. To overcome this gap, Xavier et al. (2022) applied the interpolation method (inverse distance weighting and angular distance weighting - IDW) to generate the gridded value for climate variables such as precipitation. Besides IDW, there is the spatio-temporal Kriginghe method \citep{aryaputera2015very}. However, the interpolation method has limitations as the error is proportional to the distance between the data points to the power n \citep{emetere2022numerical}. To minimize it, \cite{stauffer2017spatio} proposed a spatio-temporal model, using a left-censored normal distribution approach. Several methods are employed in different fields to estimate spatio-temporal data. For instance, \cite{blangiardo2013spatial} proposed the INLA approach for epidemiological data, while \cite{gamerman2022dynamic} suggested reducing spatio-temporal dimensionality using dynamic structural equations to analyze the effects of air pollution on hospitalization data. Focusing on climatological data, \cite{daly2008physiographically} and \cite{thornton1997generating} explored climate patterns using regional regression models. \cite{lewis2006error} utilized generalized additive models, emphasizing their flexibility to represent cyclical temporal effects and spatial distribution. In a Bayesian context, \cite{laurini2019spatio} proposed a spatio-temporal model to analyze climate change in temperature series, incorporating fixed effects, trends, seasonality, dynamic cycles, and a spatio-temporal random component.

Functional components based on space and time are used as linear predictors due to the scarcity of information at a geographic location, which renders techniques like regional regression unfeasible \citep{daly2008physiographically}, as well as their ability to explain nonlinear relationships between the variable of interest and space-time conditions, as highlighted by \cite{guan2009modeling}. Additionally, the potential for a nonlinear relationship between the variable of interest and space-time conditions, as highlighted by \cite{guan2009modeling}, serves as another justification. Lastly, the inclusion of functional random effects allows for the incorporation of spatial and temporal correlations, expanding the model’s capabilities beyond those considered in the random error \citep{stauffer2017spatio}.

This work aims to develop a spatio-temporal regression model with a block structure, incorporating fixed and random functional variables as predictors for the response variable, using the Functional Data Analysis (FDA) approach \citep{ReS2002}. Each observation is modeled by fixed and random spatio-temporal effects, which are approximated by linear combinations of tensor product of B-spline bases evaluated in time and space \citep{Boor1978}. The covariance structures considered account for spatio-temporal correlations among measurements within the same block and repetition. The expansion in cubic B-splines accommodates intra-block spatial and temporal correlations. The block structure of the model is designed to capture seasonality in precipitation data \citep{laurini2019spatio}. Parameter estimation is carried out using a Bayesian approach, as described by \citep{laurini2019spatio}.

Next, we describe the structure of this work. In Section \ref{sect:projection}, we present the proposed model, which considers a model with spatio-temporal response and functional covariates. In Section \ref{sect:bayesian}, we develop Bayesian inference for the proposed model. In Section \ref{sect:simulation}, we present the results of an extensive set of simulations aimed at evaluating the performance of the proposed model from the following perspectives: sensitivity of the prior distribution of variance components, effect of sample size and the number of repeated measures on the estimation process of the model parameters, determination of the parameter $\kappa$, which defines the order of the modified Bessel function of the third kind in the Matérn correlation structure \citep{Matern1999}, and, finally, the predictive capability of the proposed model. In Section \ref{dados_reais}, we present the estimation and prediction results for the proposed model using precipitation data from the state of Goiás, Brazil, collected between 1980 and 2001. In Section \ref{conclusao}, we provide the final considerations regarding the proposed model.

\section{The Spatial-Temporal Functional Model for Data with Block Structure and Repeated Measures}\label{sect:projection}

	This study proposes a spatial-temporal model that incorporates both fixed and random effects curves as predictors while accommodating a repeated measures block structure. Let $Y_{ij}(\mathbf{x},t)$ be the repeated measure $j = 1,\ldots, J_i$ within block $i = 1,\ldots, I$, observed at time $t$ within the set $D_{T}\subset \mathbb{N}$ and at the location $\mathbf{x} =(x_{1}, x_{2}) \in D_{lat} \times D_{long}\subset \mathbb{R}^2$. In our context, $x_{1} \in D_{lat} $ corresponds to latitude, and $x_{2} \in D_{long}$ corresponds to longitude, both expressed in decimal degrees. The proposed model is formulated as
		
		\begin{eqnarray} \label{modelo_proposto}
			Y_{ij}(\mathbf{x},t)&=&\mu(\mathbf{x},t)+\zeta_{i}(\mathbf{x})+\Gamma_{i}(t)+ \epsilon_{ij}(\mathbf{x},t),
		\end{eqnarray} 
		
		\noindent where the processes $\epsilon_{ij}(\mathbf{x},t)$ represents the spatial-temporal random errors in the model, $\mu(\mathbf{x},t)$ represents the overall mean function at location $\mathbf{x}$ and time $t$, $\zeta_{i}(\mathbf{x})$ represents the random spatial effect associated with location $\mathbf{x}$ in block $i$, and $\Gamma_{i}(t)$ accommodates the random temporal effect associated with time $t$ and block $i$.

		In this methodology, the mean $\mu(\mathbf{x},t)$ is exclusively based on the spatio-temporal coordinates $\mathbf{x}$ and $t$. However, it is feasible to introduce extra explanatory factors into $\mu(.,.)$ by means of a linear or nonlinear model, if the covariates of interest are available for all spatio-temporal points $\mathbf{x}$ and time instance $t$. We assume that $\mu(\mathbf{x},t)$ falls within the realm of smooth functions, being represented by

		\begin{eqnarray}\label{media_modelo_proposto}	\mu(\mathbf{x},t)&=&\displaystyle\sum_{a=1}^{K_{\mu_{1}}}\displaystyle\sum_{s=1}^{K_{\mu_{2}}}\displaystyle\sum_{d=1}^{K_{\mu_{3}}}\beta_{asd}M^{(\Upsilon_{\mu_1})}_{a}(x_{1})M^{(\Upsilon_{\mu_2})}_{s}(x_{2})M^{(\Upsilon_{\mu_3})}_{d}(t),
		\end{eqnarray}
		
		\noindent where $K_{\mu_{1}}$, $K_{\mu_{2}}$, and $K_{\mu_{3}}$ are positive integers representing the number of bases related to the latitude, longitude, and time spaces, respectively, with $K_{\mu_j}-4$ interior nodes taken according to the sets $\Upsilon_{\mu_j}$, for $j=1,2,3$, where $\Upsilon_{\mu_{1}} \subset D_{lat}$, $\Upsilon_{\mu_{2}} \subset D_{long}$, and $\Upsilon_{\mu_{3}} \subset D_{T}$. Additionally, $M^{(\Upsilon_{\mu_{1}})}_{a}(.)$, $M^{(\Upsilon{\mu_2})}_{s}(.)$, and $M^{(\Upsilon{\mu_3})}_{d}(.)$ are cubic B-splines bases evaluated on $D_{lat}$, $D_{long}$, and $D_{T}$, respectively, and $\beta_{asd}$ is the coefficient in the linear combination of the B-splines basis expansion, representing the components of the vector $\mbox{\boldmath $\beta$}$.
		
		For the spatial component $\zeta_{i}(\mathbf{x})$, consider the following tensor product of cubic B-splines bases evaluated on $D_{lat}$ and $D_{long}$
  
	\begin{eqnarray}\label{efeito_espacial_modelo_proposto}
\zeta_{i}(\mathbf{x}) &=& \displaystyle\sum_{f=1}^{K_{\zeta_{1}}}\displaystyle\sum_{g=1}^{K_{\zeta_{2}}}\theta_{fg}^{(i)}M^{(\Upsilon_{\zeta_{1}})}_{f}(x_{1})M^{(\Upsilon_{\zeta_2})}_{g}(x_{2}),
		 \end{eqnarray}
		 
\noindent where $M^{(\Upsilon_{\zeta_1})}_{f}(.)$ and $M^{(\Upsilon_{\zeta_2})}_{g}(.)$ denote the cubic B-spline bases, evaluated respectively over the domains $D_{lat}$ and $D_{long}$. Specifically, $K_{\zeta_1}$ and $K_{\zeta_2}$ represent the number of bases associated with the latitude and longitude dimensions. Within these, the $K_{\zeta_j}-4$ interior nodes are defined within the sets $\Upsilon_{\zeta_j}$ for $j \in \{1,2\}$. It is worth noting that $\Upsilon_{\zeta_1}$ is a subset of $D_{lat}$, while $\Upsilon_{\zeta_2}$ is a subset of $D_{long}$. Additionally, $\theta_{fg}^{(i)}$ is a random coefficient in the linear combination of the B-splines expansion of the spatial component of block $i$ and represents the components of the vector ${\bf{\Theta}}^{(i)}$, which follows a normal distribution with mean $\mathbf{0}$ and covariance matrix ${\bf{\Sigma_{\Theta}}}^{(i)}$. Finally,
        
        \begin{eqnarray}\label{efeito_temporal_modelo_proposto}
		 	\Gamma_{i}(t) &=& b\displaystyle\sum_{l=1}^{K_{\Gamma}}\vartheta_{l}^{(i)}M^{(\Upsilon_{\Gamma})}_{l}(t),
		 \end{eqnarray}
		 
\noindent in which $M^{(\Upsilon_{\Gamma})}_{l}(.)$ are the $K_{\Gamma}$ cubic B-splines bases evaluated on $D_{T}$, with $\Upsilon_{\Gamma} \subset D_{T}$ representing the set of $K_{\Gamma}-4$ interior nodes, and $\vartheta_{l}^{(i)}$ is a random coefficient of the linear combination of the B-splines expansion of the temporal component of block $i$ and represents the components of the vector $\mbox{\boldmath $\vartheta$}^{(i)}$, which follows a normal distribution with mean $\mathbf{0}$ and covariance matrix $\bf{\Sigma_{\mbox{\boldmath $\vartheta$}}}^{(i)}$. Further assume that $\mathbf{\Theta}^{(i)}$ and $\mbox{\boldmath $\vartheta$}^{(i)}$ are independent random variables for $i=1,\ldots,I$, whose respective covariance structures accommodate the correlations between the repeated measurements in block $i$, as observed through the B-splines bases.

        We define $\mathbf{Y}$ as $(\mathbf{Y}_{1}^{\top},\mathbf{Y}_{2}^{\top},\ldots,
 \mathbf{Y}_{I}^{\top})^{\top}$ through the regression
		  
    \begin{eqnarray} \label{y_vetorizada}
		  	\mathbf{Y}&=&\mathbf{X}\mbox{\boldmath $\beta$}+\mathbf{Q}\mathbf{\Theta}+\mathbf{R}\mbox{\boldmath $\vartheta$}+\mathbf{E},
     \end{eqnarray} 
		  
  \noindent where $\mathbf{X}=((\mathbf{1}_{J_1} \otimes \mathbf{M})^{\top},(\mathbf{1}_{J_2} \otimes \mathbf{M})^{\top},\ldots,(\mathbf{1}_{J_I} \otimes \mathbf{M})^{\top})$, $\mathbf{Q}=\mathbf{BD}(\{(\mathbf{1}_{J_i} \otimes \mathbf{P})\}_{i=1}^{I})$, $\mathbf{BD}(.)$ representing a block-diagonal matrix,  $\mathbf{R}=\mathbf{BD}(\{(\mathbf{1}_{J_i} \otimes \mathbf{N})\}_{i=1}^{I})$, $\mathbf{N}=(\mathbf{1}_{n \times 1}\otimes{\mathbf{M}^{\Gamma^{(T)}}})$, $\mathbf{\Theta}=(\mathbf{\Theta}^{(1)}, \mathbf{\Theta}^{(2)}, \ldots, \mathbf{\Theta}^{(I)})$, $\mbox{\boldmath $\vartheta$}=(\mbox{\boldmath $\vartheta$}^{(1)}, \mbox{\boldmath $\vartheta$}^{(2)}, \ldots, \mbox{\boldmath $\vartheta$}^{(I)})$, with

		  \begin{eqnarray}\label{M}
		  	\mathbf{M}&=&\left(\begin{array}{r}
		  		\mathbf{M}^{(\Upsilon_{\mu_1})}[1,] \otimes \mathbf{M}^{(\Upsilon_{\mu_2})}[1,]\otimes \mathbf{M}^{(\Upsilon_{\mu_3})}\\
		  		\mathbf{M}^{(\Upsilon_{\mu_1})}[2,] \otimes \mathbf{M}^{(\Upsilon_{\mu_2})}[2,]\otimes \mathbf{M}^{(\Upsilon_{\mu_3})}\\
		  		\vdots\\
		  		\mathbf{M}^{(\Upsilon_{\mu_1})}[n,] \otimes \mathbf{M}^{(\Upsilon_{\mu_2})}[n,]\otimes \mathbf{M}^{(\Upsilon_{\mu_3})}\\
		  	\end{array}\right),
		  \end{eqnarray}
		
		\noindent and
		  
		  \begin{eqnarray}\label{P}
		  	\mathbf{P}&=&\left(\begin{array}{r}
		  		\mathbf{1}_{\tau \times 1}\otimes \mathbf{M}^{(\Upsilon_{\zeta_{1}})}[1,]\otimes \mathbf{M}^{(\Upsilon_{\zeta_{2}})}[1,] \\
		  		\mathbf{1}_{\tau \times 1}\otimes \mathbf{M}^{(\Upsilon_{\zeta_{1}})}[1,]\otimes \mathbf{M}^{(\Upsilon_{\zeta_{2}})}[1,] \\
		  		\vdots\\
		  		\mathbf{1}_{\tau \times 1}\otimes \mathbf{M}^{(\Upsilon_{\zeta_{1}})}[n,]\otimes \mathbf{M}^{(\Upsilon_{\zeta_{2}})}[n,] \\ 
		  	\end{array}\right).
		  \end{eqnarray}

Additionally, the vector of random errors $\mathbf{E}$ follows a normal distribution with block-diagonal covariance matrix  $\mathbf{\Sigma}_{\epsilon}:= \mathbf{BD}(\{\mathbf{\Sigma}_{\epsilon_{i}}\}_{i=1}^{I})$.

	In the literature, we find a variety of covariance structures available for spatial-temporal models, including the exponential model, the Gaussian model, the Matérn model, the nugget model, and others \citep{cressie2015statistics}. In this study, we have selected $\mathbf{\Sigma_{\Theta}}^{(i)} = \sigma^{2}_{\theta_i}\mathbf{I}_{K_{\zeta}}$, $\mathbf{\Sigma_{ \mbox{\boldmath $\vartheta$}}}^{(i)} = \sigma^{2}_{\vartheta_i}\mathbf{I}_{K_{\Gamma}}$ e $\mathbf{\Sigma}_{\epsilon_{ij}} := \mathbf{\Sigma_{\epsilon}} =\omega^2 \mathbf{\Sigma_{S}}(\kappa,\phi) \otimes \mathbf{\Sigma}_{T}(\varphi)$, where $\mathbf{\Sigma}_{T}(\varphi)$ denotes a correlation structure characterized by an exponential decay pattern parameterized by $\varphi$, and $\mathbf{\Sigma}_{S}(\kappa,\phi)$ incorporates a Matérn correlation structure defined by the parameters $\kappa$ and $\phi$, defined by 
		 
	\begin{eqnarray}\label{matern}
		 	\mathbf{\Sigma}_{S}(d | \kappa,\phi) &=& \{2^{\kappa-1}\Gamma(\kappa)\}^{-1}(d/\phi)^{\kappa}B_{\kappa}(d/\phi),
		 \end{eqnarray}
   
        \noindent where $B_{\kappa}(.)$ represents the modified Bessel function \citep{abramowitz1970handbook}. Thus, the covariance structure of the proposed model can be derived as

		 \begin{eqnarray}\label{corr_geral}
		 	{\rm cov}(Y_{ij}(\mathbf{x}_r,t_s),Y_{i^{'}j^{'}}(\mathbf{x}_p,t_q))&=& \sigma_{\theta_i}\{\mathbf{M}^{(\Upsilon_{\zeta_{1}})}[r,] \otimes \mathbf{M}^{(\Upsilon_{\zeta_{2}})}[r,]\} \nonumber \\
		 	&\times& \{\mathbf{M}^{(\Upsilon_{\zeta_{1}})}[p,] \otimes \mathbf{M}^{(\Upsilon_{\zeta_{2}})}[p,]\}^{\top} \mathbf{I}_{\{i=i^{'}\}}\nonumber \\
		 	&+&  \sigma_{\vartheta_i} \mathbf{M}^{(\Upsilon_{\Gamma})}[s,]\mathbf{M}^{(\Upsilon_{\Gamma})^{\top}}[q,]\mathbf{I}_{\{i=i^{'}\}} \nonumber \\
		 	&+& \omega^2 \mathbf{\Sigma}_{S}(\mathbf{x}_r,\mathbf{x}_s \hbox{ | }\kappa,\phi) \mathbf{\Sigma}_{T}(t_s.t_q | \varphi) \mathbf{I}_{\{i=i^{'}\}}\mathbf{I}_{\{j=j^{'}\}}. 
		 \end{eqnarray}
		 	 
	Furthermore, we assumed that the sets of interior knots in all dimensions within the proposed model in Equation \eqref{modelo_proposto} were uniformly spaced along their respective dimensions. From now on we simplify the notation by considering $K_{\mu}:=K_{\mu_{1}}=K_{\mu_{2}}=K_{\mu_{3}}$ and $K_{\zeta}:=K_{\zeta_{1}}=K_{\zeta_{2}}$.

\section{Bayesian Approach} \label{sect:bayesian}

As highlighted by \cite{stein1999interpolation}, estimating the parameter $\kappa$ in the Matérn correlation structure in \eqref{matern} is not recommended due to identification challenges and resulting instability. The parameter $\kappa$ has a significant impact on the smoothness of the correlation function, influencing the covariance functions and, consequently, the behavior of the stochastic process in the context of Bayesian estimation, which can lead to imprecise estimates and unreliable results. Therefore, we opted to fix this parameter through the inference analysis. The parameter vector for the model proposed in (\ref{modelo_proposto}) is represented by $\bxi=(\bbeta,\sigma^{2}_{\theta_1},...,\sigma^{2}_{\theta_I},\sigma^{2}_{\vartheta_1},...,\sigma^{2}_{\vartheta_I},\omega^2,\phi,\varphi)$. Let $\mathbf{y}=(\mathbf{y_1},\ldots,\mathbf{y_I})$ be a sample of $\mathbf{Y}_{i}\sim \mathbf{N}(\mathbf{M}_{i} \bbeta,\sigma_{\theta_i}\mathbf{P}_{i}\mathbf{P}_{i}^{\top}+\sigma_{\vartheta_i}\mathbf{N}_{i}\mathbf{N}_{i}^{\top} + \mathbf{\Sigma}_{\epsilon_{i}})$. The conditional likelihood function is given by

\begin{eqnarray}\label{verrossi_geral}
 L(\mbox{\boldmath $\xi$}|\mathbf{y},\mathbf{\Theta},\mbox{\boldmath $\vartheta$}) &=& \prod_{i=1}^{I}p(\mathbf{y}_{i}|\mbox{\boldmath $\xi$},\mathbf{\Theta}^{(i)},\mbox{\boldmath $\vartheta$}^{(i)}) \mbox{.}
\end{eqnarray} 
	
Let $\Pi(\mbox{\boldmath $\xi$}|\mathbf{\Delta})$ be the joint prior distribution of $\bxi$, where $\bDelta$ is a known vector of hyperparameters. Consequently, the joint posterior distribution of $\bxi$, augmented by the random vectors $\bTheta$ and $\bvartheta$, can be expressed as
	
	\begin{eqnarray}\label{posteriori}
		p(\mbox{\boldmath $\xi$},\mathbf{\Theta},\mbox{\boldmath $\vartheta$}|\mathbf{y},\mathbf{\Delta}) & \propto & p(\mathbf{y}|\mbox{\boldmath $\xi$},\mathbf{\Theta},\mbox{\boldmath $\vartheta$})p(\mathbf{\Theta}|\sigma^{2}_{\theta_i})p(\mbox{\boldmath $\vartheta$}|\sigma^{2}_{\vartheta_i})\Pi(\mbox{\boldmath $\xi$}|\mathbf{\Delta}) \nonumber \\
		&=&\prod_{i=1}^{I}\prod_{j=1}^{J_i}p(\mathbf{y}_{ij}|\mbox{\boldmath $\xi$},\mathbf{\Theta}^{(i)},\mbox{\boldmath $\vartheta$}^{(i)})p(\mathbf{\Theta}^{(i)}|\sigma^{2}_{\theta_i})p(\mbox{\boldmath $\vartheta$}^{(i)}| \sigma^{2}_{\vartheta_i})\Pi(\mbox{\boldmath $\xi$}|\mathbf{\Delta}), 
	\end{eqnarray}
	
	\noindent where $p(\mathbf{y}_{ij}|\mbox{\boldmath $\xi$},\mathbf{\Theta}^{(i)},\mbox{\boldmath $\vartheta$}^{(i)})$ represents the probability density of a multivariate normal distribution with a mean of $\mathbf{M}\mbox{\boldmath $\beta$}+ \mathbf{P}\mathbf{\Theta}^{(i)}+ \mathbf{N}\mbox{\boldmath $\vartheta$}^{(i)}$ and a covariance matrix of $\omega^2 \mathbf{\Sigma}_{S}(\kappa,\phi) \otimes \mathbf{\Sigma}_{T}(\varphi)$. Furthermore, $p(\mathbf{\Theta}^{(i)}|\sigma^{2}_{\theta_i})$ denotes the probability density of a normal distribution with a mean of zero and a covariance matrix of $\sigma^{2}_{\theta_i}\mathbf{I}_{K_{\zeta}^2}$, and $p(\mbox{\boldmath $\vartheta$}^{(i)}|\sigma^{2}_{\vartheta_i})$ is the probability density of a normal distribution with a mean of zero and a covariance matrix of $\sigma^{2}_{\vartheta_i}\mathbf{I}_{K_{\Gamma}^2}$.

    In this study, we assume that  the elements of the vector parameter $\bxi$ are independent and can be selected from the distribution families listed in Table \ref{cond_completas}.
    
    To assign values for the parameter $\kappa$, we applied two established and widely recognized criteria from the literature. The first criterion relies on the logarithm of the marginal pseudo-likelihood (LPML), which is derived from the conditional predictive ordinate introduced by \cite{Geisser1979} and further developed in \cite{Geisser1993}. The second criterion is a modified version of the deviance information criterion (DIC) tailored specifically for situations involving random effects, as proposed by \cite{Celeux2006}.

Due to the complexity of the proposed model, the joint augmented posterior distribution of $(\boldsymbol{\xi}, \mathbf{\Theta}, \boldsymbol{\vartheta})$, as outlined in equation \eqref{posteriori}, does not have a closed-form. Under this circumstance, we employ the Gibbs sampler procedure \citep{robert1999monte} to generate samples from the joint augmented posterior distribution. As it is well known, the Gibbs sampler relies on full conditional distributions, which often have known closed-forms \citep{Gamerman2006}. In cases where the full conditional distribution does not have closed forms, a step of the Metropolis-Hastings algorithm is integrated into the Gibbs sampler scheme, as discussed by \citep{Gamerman2006}.

The full conditional distributions are outlined in Table \ref{cond_completas}. It is evident that the full conditional distributions for the parameters $\varphi$ and $\phi$ are not readily available for direct sampling. Therefore, the Metropolis-Hastings algorithm was considered for sampling from these distributions. These full conditional distributions, as summarized in Table \ref{cond_completas}, consider

	\begin{eqnarray}
		\tilde{\mbox{\boldmath $\beta$}} &:=& [\mathbf{X}^{\top}\mathbf{\Sigma}_{\epsilon}^{-1}\mathbf{X}+\omega^2/\sigma^{2}_{\beta}\mathbf{I}]^{-1}\mathbf{X}^{\top}\mathbf{\Sigma}_{\epsilon}^{-1}(\mathbf{Y-Q\mathbf{\Theta}-R\mbox{\boldmath $\vartheta$}})
	\end{eqnarray}

	\begin{eqnarray}
		\mathbf{\tilde{\Theta}}^{(i)}&:=&[\mathbf{P}_{i}^{\top}\mathbf{\Sigma}_{\epsilon_i}^{-1}\mathbf{P}_{i}+\omega^2/\sigma^{2}_{\theta_i}\mathbf{I}]^{-1}\mathbf{P}_{i}^{\top}\mathbf{\Sigma}_{\epsilon_i}^{-1}(\mathbf{Y}_{i} -\mathbf{M}_{i}\mbox{\boldmath{$\beta$}}-\mathbf{N}_{i}\mbox{\boldmath{$\vartheta$}}^{(i)}),
	\end{eqnarray}

	\begin{eqnarray}
		\tilde{\mbox{\boldmath $\vartheta$}}^{(i)}&:=&[\mathbf{N}_{i}^{\top}\mathbf{\Sigma}_{\epsilon_i}^{-1}\mathbf{N}_{i}+\omega^2/\sigma^{2}_{\vartheta_i}\mathbf{I}]^{-1} \mathbf{N}_{i}^{\top}\mathbf{\Sigma}_{\epsilon_i}^{-1}(\mathbf{Y}_{i} -\mathbf{M}_{i}\mbox{\boldmath{$\beta$}}-\mathbf{P}_{i}\mathbf{\tilde{\Theta}}^{(i)}),
	\end{eqnarray}

	\noindent e

	\begin{eqnarray}
		\tilde{s}^{2}_{\omega^2}&:=&(\mathbf{Y}-\mathbf{X}\mbox{\boldmath{$\beta$}}-\mathbf{Q}\mathbf{\Theta}-\mathbf{R}\mbox{\boldmath{ $\vartheta$}})^{\top}\mathbf{\Sigma}^{-1}_{\epsilon}(\mathbf{Y}-\mathbf{X}\mbox{\boldmath{$\beta$}}-\mathbf{Q}\mathbf{\Theta}-\mathbf{R}\mbox{\boldmath{$\vartheta$}}).
	\end{eqnarray}

\begin{sidewaystable}
\caption{Proposed prior distributions and their respective full conditional distributions for the elements in $\boldsymbol{\xi}$. \label{cond_completas}}
\centering
\begin{tabular*}{\textheight}{@{\extracolsep\fill}lcc@{\extracolsep\fill}}
\toprule
\textbf{Parameter} & \textbf{Prior distribution} & \textbf{Full conditional distribution} \\
\midrule
$\mbox{\boldmath $\beta$}$ & $\mathbf{N}(\mathbf{m_{\beta}},\sigma^{2}_{\beta} \mathbf{I})$ & $\mathbf{N}[\tilde{\mbox{\boldmath $\beta$}},\omega^2(\mathbf{X}^{\top}\mathbf{\Sigma}_{\epsilon}^{-1}\mathbf{X}+\omega^2/\sigma^{2}_{\beta}\mathbf{I})]$ \\
$\mathbf{\Theta}^{(i)}$ & $\mathbf{N}(\mathbf{0},\sigma^{2}_{\theta_i} \mathbf{I})$ & $\mathbf{N}[\mathbf{\tilde{\Theta}}^{(i)},\omega^2(\mathbf{P}_{i}^{\top}\mathbf{\Sigma}_{\epsilon_{i}}^{-1}\mathbf{P}_{i}+\omega^2/\sigma^{2}_{\theta_i}\mathbf{I})]$ \\
$\mbox{\boldmath $\vartheta$}^{(i)}$ & $\mathbf{N}(\mathbf{0},\sigma^{2}_{\vartheta_i} \mathbf{I})$ & $\mathbf{N}[\tilde{\mbox{\boldmath $\vartheta$}}^{(i)},\omega^2(\mathbf{N}_{i}^{\top}\mathbf{\Sigma}_{\epsilon_{i}}^{-1}\mathbf{N}_{i}+\omega^2/\sigma^{2}_{\vartheta_i}\mathbf{I})]$ \\
$\sigma^{2}_{\theta_i}$ & $\hbox{Scaled Inverse Chi-Square}(v_{\theta_i},s^{2}_{\theta_i})$ & $\hbox{Scaled Inverse Chi-Square}\left(K_{\zeta}^{2}+v_{\theta_i},\dfrac{\mathbf{\Theta}^{(i)^{\top}}\mathbf{\Theta}^{(i)}+v_{\theta_i}s^{2}_{\theta_i}}{K_{\zeta}^{2}+v_{\theta_i}}\right)$ \\
$\sigma^{2}_{\theta_i}$ & $\hbox{Inverse Gamma}(a_{\theta_i},b_{\theta_i})$ & $\hbox{Inverse Gamma }\left(\dfrac{K_{\zeta}^{2}}{2}+a_{\theta_i},\dfrac{\mathbf{\Theta}^{(i)^{\top}}\mathbf{\Theta}^{(i)}}{2}+b_{\theta_i}\right)$ \\
$\sigma^{2}_{\vartheta_i}$ & $\hbox{Scaled Inverse Chi-Square}(v_{\vartheta_i},s^{2}_{\vartheta_i})$ & $\hbox{Scaled Inverse Chi-Square}\left(K_\Gamma+ v_{\vartheta_i},\dfrac{\mbox{\boldmath $\vartheta$}^{(i)^{\top}}\mbox{\boldmath $\vartheta$}^{(i)}+v_{\vartheta_i}s^{2}_{\vartheta_i}}{K_\Gamma+v_{\vartheta_i}}\right)$ \\
$\sigma^{2}_{\vartheta_i}$ & $\hbox{Inverse Gamma}(a_{\vartheta_i},b_{\vartheta_i})$ & $\hbox{Inverse Gamma}\left(\dfrac{K_\Gamma}{2}+a_{\theta_i},\dfrac{\mbox{\boldmath $\vartheta$}^{(i)^{\top}}\mbox{\boldmath $\vartheta$}^{(i)}}{2}+b_{\vartheta_i}\right)$ \\
$\omega^{2}$ & $\hbox{Scaled Inverse Chi-Square}(v_{\omega^{2}},s^{2}_{\omega^{2}})$ & $\hbox{Scaled Inverse Chi-Square} \left[n\tau(J_1+J_2+J_3)+v_{\omega^{2}},\dfrac{\tilde{s}^{2}_{\omega^2}+v_{\omega^2}s^{2}_{\omega^{2}}}{n\tau(J_1+J_2+J_3)+v_{\omega^{2}}}\right]$ \\
$\omega^{2}$ & $\hbox{Inverse Gamma}(a_{\omega^{2}},b_{\omega^{2}})$ & $\hbox{Inverse Gamma }\left(\dfrac{n\tau(J_1+J_2+J_3)}{2}+a_{\omega^{2}},\frac{\tilde{s}^{2}_{\omega^2}}{2}+b_{\omega^{2}}\right)$ \\
$\phi$ & $\hbox{Gamma}(c_\phi,d_\phi)$ & $p(\mathbf{y}|\mbox{\boldmath $\xi$},\mathbf{\Theta},\mbox{\boldmath $\vartheta$})p(\phi|c_\phi,d_\phi)$ \\
$\varphi$ & $\hbox{Gamma}(c_\varphi,d_\varphi)$ & $p(\mathbf{y}|\mbox{\boldmath $\xi$},\mathbf{\Theta},\mbox{\boldmath $\vartheta$})p(\varphi|c_\varphi,d_\varphi)$ \\
\bottomrule
\end{tabular*}
\begin{tablenotes}
\end{tablenotes}
\end{sidewaystable}

\newpage
\section{Simulation Study}\label{sect:simulation}

We carried out a simulation study to evaluate the effectiveness of the model presented in (\ref{modelo_proposto}) from several perspectives. In the subsequent sections, we explore:

\begin{enumerate}
\item The sensitivity of the prior distribution of the variance components (Subsection \ref{sim1});
\item The influence of sample size and the number of repeated measures on parameter estimation (Subsection \ref{sim2});
\item The procedure for selecting the parameter $\kappa$ (Subsection \ref{sim3});
\item The model's predictive capabilities (Subsection \ref{sim_4}).
\end{enumerate}

\subsection{Sensitivity Analysis of the Prior Distribution of Variance Components} \label{sim1}

Our objective in this simulation study is to evaluate the influence of different prior distributions on the estimation of variance components. Considering the model proposed in \ref{modelo_proposto} and the set of parameters presented in Table \ref{reais}, with $\kappa=0.2$ fixed and values of $K_{\mu}=9$, $K_{\zeta}=5$, and $K_{\Gamma}=7$, we generated 50 independent datasets. Each dataset contains three blocks, with $J_1=9$ and $J_2=J_3=6$ repeated measurements, respectively, resulting in a total of $\tau=6$ observed time points, thus mirroring the structure observed in the real dataset detailed in Section \ref{dados_reais}.

	\begin{table}[t]
	    	\centering
	    	\caption{Proposed values for variance parameters for the model in Equation (\ref{modelo_proposto}) used to generate 50 independent datasets, fixing $\kappa=0.2$, $K_{\mu}=9$, $K_{\zeta}=5$ e $K_{\Gamma}=7$.}
	    	\begin{tabular}{ccc}
	    		\hline
	    		Parameters & Proposed values\\ \hline
	    		$\omega^2$ & 11.02\\ 
	    		$\sigma^{2}_{\theta_1}$ & 0.09\\ 
	    		$\sigma^{2}_{\theta_2}$&  0.06\\ 
	    		$\sigma^{2}_{\theta_3}$ & 0.10\\ 
	    		$\sigma^{2}_{\vartheta_1}$ & 0.62\\ 
	    		$\sigma^{2}_{\vartheta_2}$& 0.18\\ 
	    		$\sigma^{2}_{\vartheta_3}$ & 1.05\\ 
	    		$\varphi$ & 2.04 \\ 
	    		$\phi$& 705 \\\hline
	    	\end{tabular}
	    	\label{reais}
	    \end{table}
	    
\begin{figure}[t]
\centerline{\includegraphics[width=510pt,height=20pc]{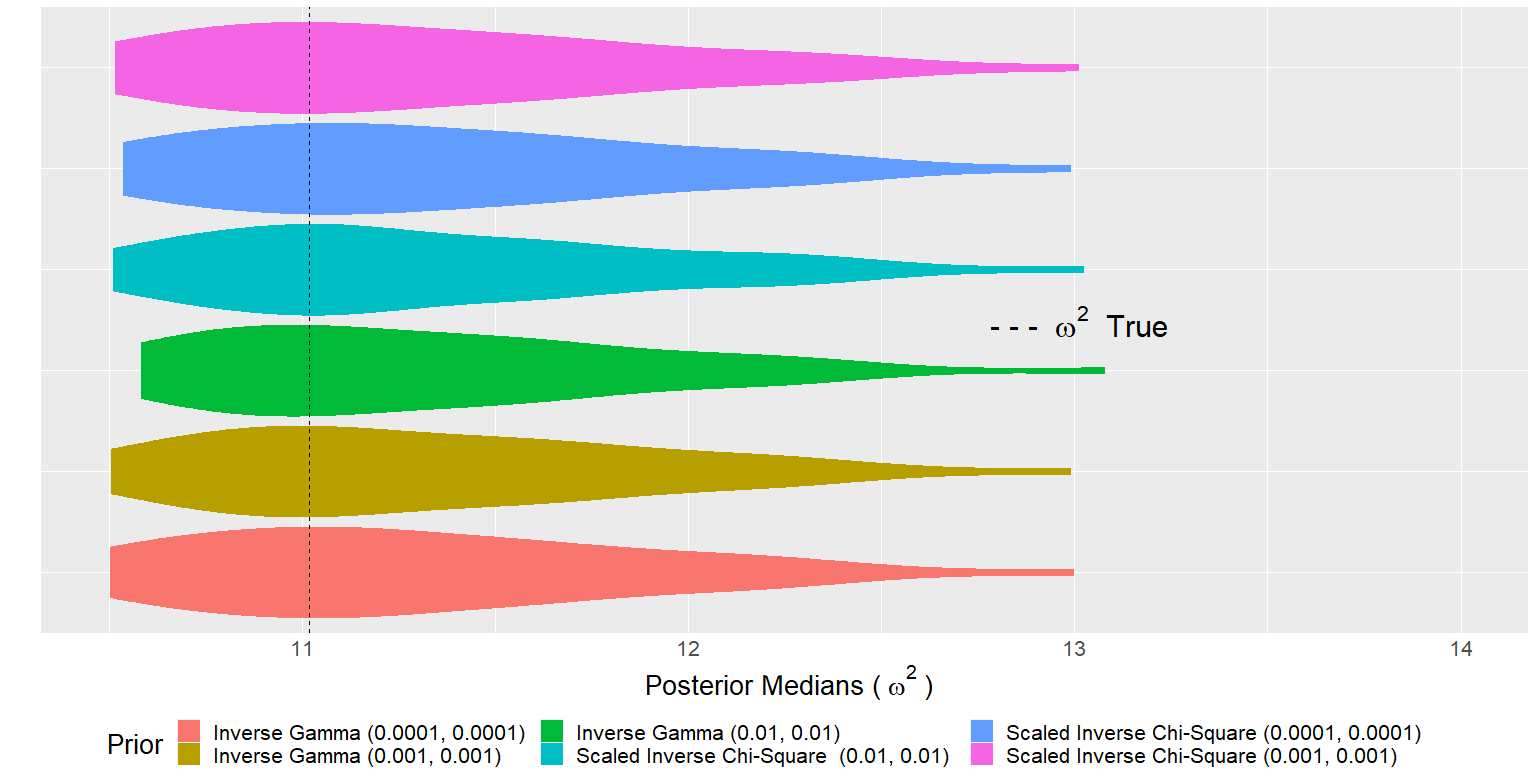}}
\caption{Posterior medians of the variance components $\omega^2$, $\varphi$, and $\phi$ of the model in (\ref{modelo_proposto}), taken across the 50 simulated datasets considering the families of Inverse Gamma and Scaled Inverse Chi-Square priors for the parameters $\omega^2$, $\sigma^{2}_{\theta_i}$, and $\sigma^{2}_{\vartheta_i}$, $i=1,2,3$, and fixing Gamma$(1,0.01)$ and Gamma$(1,0.001)$ priors for $\varphi$ and $\phi$, respectively. \label{violin_omega_phi_varphi_prioris}}
\end{figure}

\begin{figure}[t]
\centerline{\includegraphics[width=510pt,height=18pc]{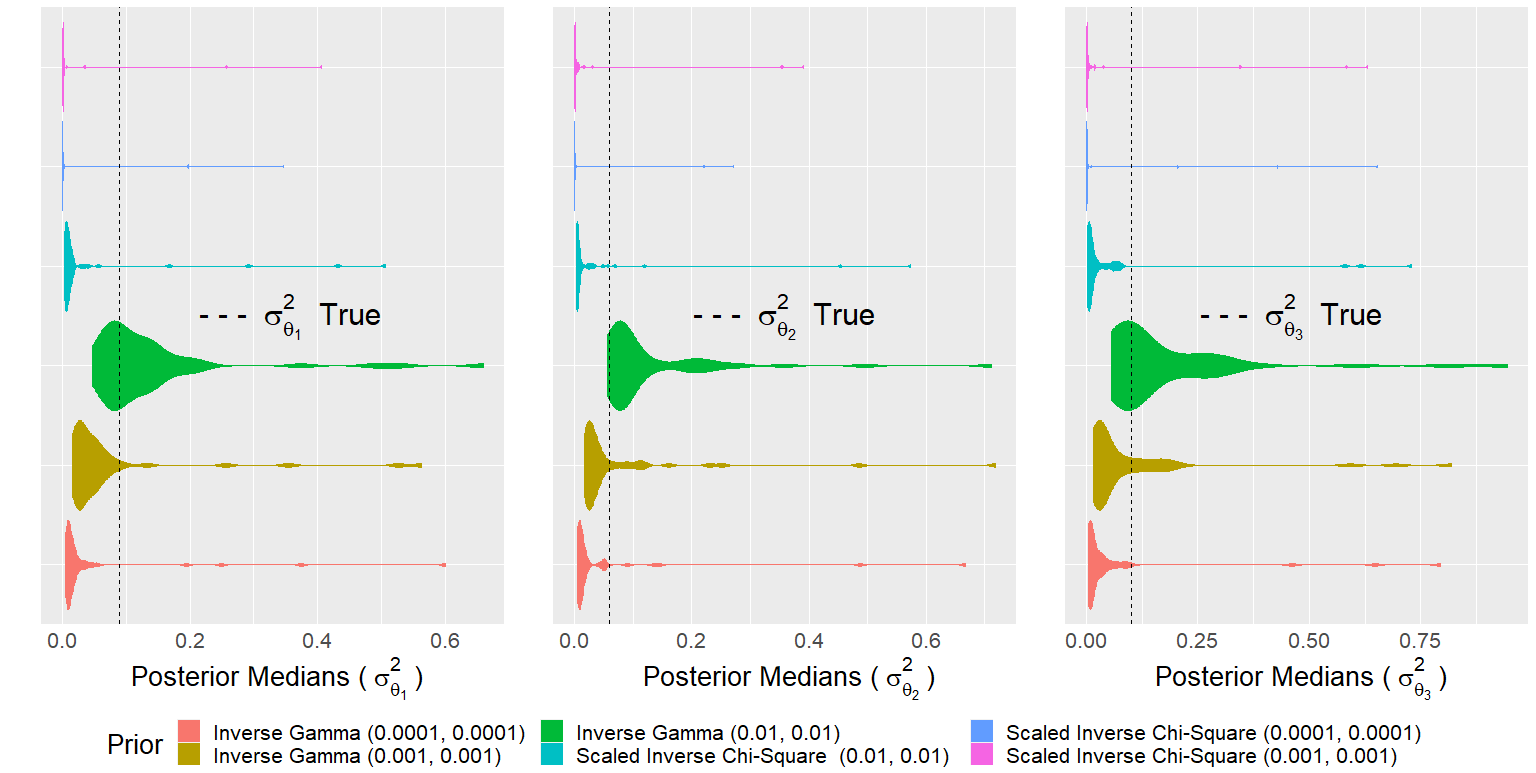}}
	    	\caption{Posterior medians of the variance components of the spatial random effect $\sigma^{2}_{\theta_i}$, $i=1,2,3$, of the model in (\ref{modelo_proposto}), taken across the 50 simulated datasets when considering the families of Inverse Gamma and Scaled Inverse Chi-Square priors for the parameters $\omega^2$, $\sigma^{2}_{\theta_i}$, and $\sigma^{2}_{\vartheta_i}$, $i=1,2,3$, and fixing Gamma$(1,0.001)$ and Gamma$(1,0.01)$ priors for the parameters $\varphi$ and $\phi$, respectively. \label{sigma2_theta_prioris}}
\end{figure}

\begin{figure}[t]
\centerline{\includegraphics[width=510pt,height=18pc]{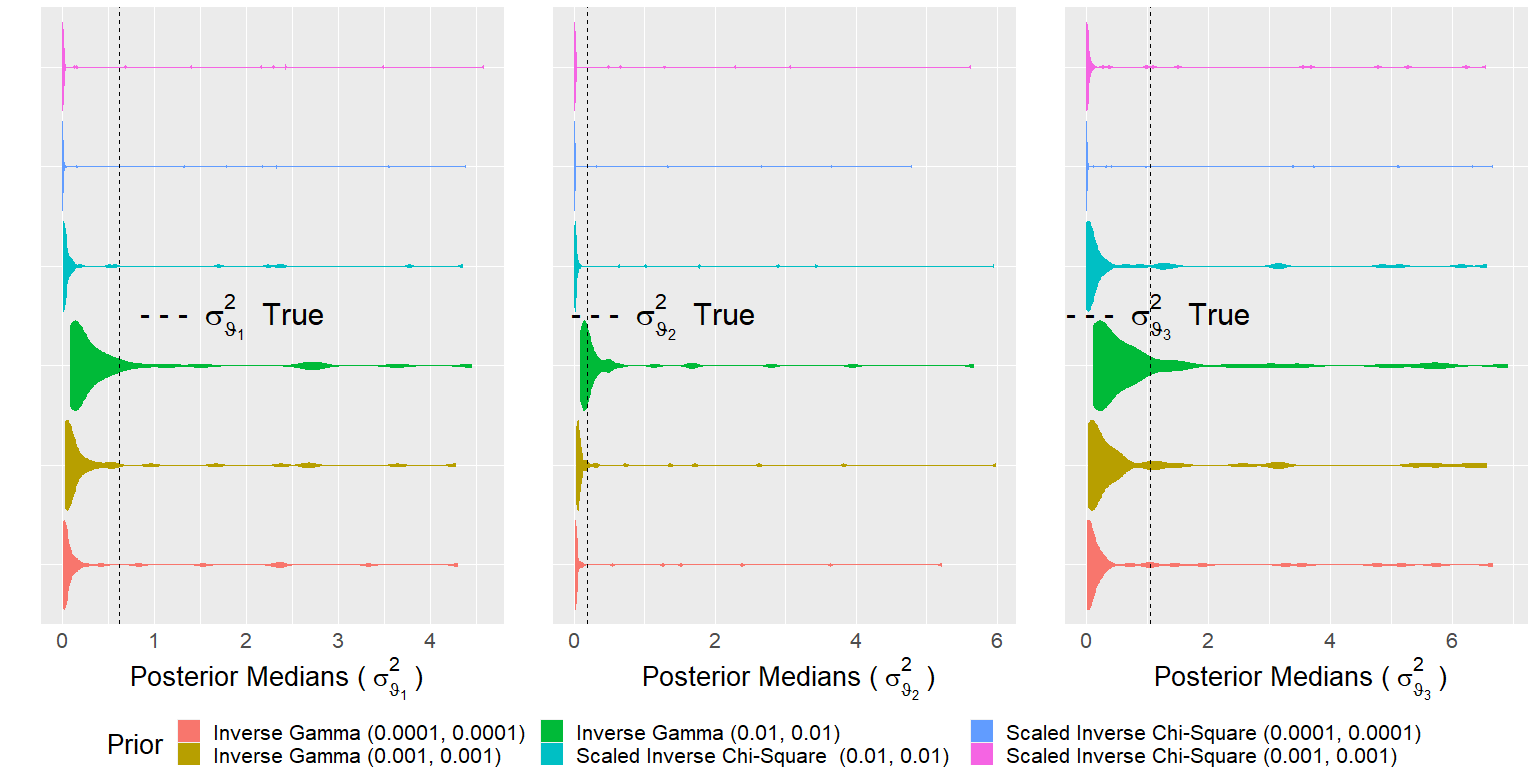}}
	    	\caption{Posterior medians of the variance components of the temporal random effect $\sigma^{2}{\vartheta_i}$, $i=1,2,3$, of the model in (\ref{modelo_proposto}), taken across the 50 simulated datasets when considering the families of Inverse Gamma and Scaled Inverse Chi-Square priors for the parameters $\omega^2$, $\sigma^{2}_{\theta_i}$, and $\sigma^{2}_{\vartheta_i}$, $i=1,2,3$, and fixing Gamma$(1,0.001)$ and Gamma$(1,0.01)$ priors for the parameters $\varphi$ and $\phi$, respectively. \label{sigma2_vartheta_prioris}}
\end{figure}

Pilot simulation studies have indicated that the use of Gamma distributions as priors for the parameters $\varphi$ and $\phi$ in the proposed model leads to quicker convergence of the chains. Consequently, in subsequent investigations, we opted to set the priors for $\varphi$ and $\phi$ as Gamma distributions with a shape parameter of $1$ and a scale parameter of $0.01$ for $\varphi$, and a shape parameter of $1$ and a scale parameter of $0.001$ for $\phi$. Features of the posterior distribution of the variance components in Table \ref{reais} are given in Figures \ref{violin_omega_phi_varphi_prioris}, \ref{sigma2_theta_prioris} and \ref{sigma2_vartheta_prioris}.

In Figure \ref{violin_omega_phi_varphi_prioris}, we present the posterior median of the variance component $\omega^2$ of model (\ref{modelo_proposto}), derived from the $50$ simulated datasets. In this study, we considered two families of priors for the parameters $\sigma^{2}_{\theta_i}$ and $\sigma^{2}_{\vartheta_i}$, using the Gamma distribution with shape and scale parameters of $1$ and $0.001$, respectively, for $\varphi$, and the Gamma distribution with a shape parameter of $1$ and a scale parameter of $0.001$ for $\phi$.  Notably, the posterior estimates for $\omega^2$ exhibit stability irrespective of changes in the priors of the other parameters.

With respect to the posterior medians of the variance components of the spatial and temporal random effects, denoted as $\sigma^{2}_{\theta_i}$ and $\sigma^{2}_{\vartheta_i}$, $i=1,2,3$, respectively, we observe in Figures \ref{sigma2_theta_prioris} and \ref{sigma2_vartheta_prioris} that the Scaled Inverse Chi-Square family tends to produce estimates near zero. More robust performance was achieved with the Inverse Gamma prior distribution, with a shape parameter of $0.01$ and a scale parameter of $0.01$, resulting in estimates that are closer to the true parameter values.

Regarding the coverage probabilities, as detailed in Tables \ref{sim6} and \ref{sim7}, it can be noted that variations in the prior distributions have a relatively modest influence on the examined parameters. The less satisfactory outcomes arise when the Scaled Inverse Chi-Square prior family is employed, whereas, overall, the most favorable outcomes are obtained when employing the Inverse Gamma prior with a shape parameter of $0.001$ and a scale parameter of $0.001$.

In summary, based on the findings presented in this section, the Inverse Gamma prior, with a shape parameter of $0.001$ and a scale parameter of $0.01$ with a shape parameter of $0.01$, emerges as the preferred choice for the parameters $\omega^2$, $\sigma^{2}_{\theta_i}$, and $\sigma^{2}_{\vartheta_i}$, where $i=1,2,3$. To ensure the convergence of the Gibbs algorithm and to guarantee uncorrelated samples, we considered a burn-in period of $120,000$ iterations and sampled every 100 iterations. Additionally, we used the Coda package \citep{CODA} in the R programming environment \citep{R}, which provides tools for diagnosing the convergence of the chains, assessing sample autocorrelation, and calculating descriptive statistics and credibility intervals. This ensures the robustness and accuracy of the conducted inferences.

 \begin{table}[t]
	    	\centering
	    	\caption{Probability of coverage (PC) for the parameters of the proposed model in (\ref{modelo_proposto}) taken over the 50 independently generated datasets using the Inverse Gamma priors for the parameters $\omega^2$, $\sigma^{2}{\theta_i}$, and $\sigma^{2}{\vartheta_i}$, $i=1,2,3$, and fixing Gamma$(1,0.01)$ and Gamma$(1,0.001)$ priors for the parameters $\varphi$ and $\phi$, respectively.}
	    	\begin{tabular}{cc|c}
	    		\hline
	    		Prior Distribution & Variance Component & \multicolumn{1}{c}{PC} \\ \hline
	    		\multirow{6}{*}{$\hbox{Inverse Gamma }(0.0001,0.0001)$}          &$\omega^2$               &$92\% $\\
	    		&$\sigma^{2}_{\theta_1}$   &$100\%$\\
	    		&$\sigma^{2}_{\theta_2}$   &$100\%$\\
	    		&$\sigma^{2}_{\theta_3}$   &$100\%$\\
	    		&$\sigma^{2}_{\vartheta_1}$&$100\%$\\
	    		&$\sigma^{2}_{\vartheta_2}$&$100\%$\\
	    		&$\sigma^{2}_{\vartheta_3}$&$96\%$\\\hline
	    		\multirow{1}{*}{$\hbox{Gamma}(1,0.01)$}      
	    		&$\varphi$                 &$90\%$\\
	    		\multirow{1}{*}{$\hbox{Gamma}(1,0.001)$}      
	    		&$\phi$                 &$92\%$\\\hline\hline
	    		\multirow{6}{*}{$\hbox{Inverse Gamma}(0.001,0.001)$}            &$\omega^2$               &$94\%$\\
	    		&$\sigma^{2}_{\theta_1}$   &$100\%$\\
	    		&$\sigma^{2}_{\theta_2}$   &$100\%$\\
	    		&$\sigma^{2}_{\theta_3}$   &$100\%$\\
	    		&$\sigma^{2}_{\vartheta_1}$&$100\%$\\
	    		&$\sigma^{2}_{\vartheta_2}$&$100\%$\\
	    		&$\sigma^{2}_{\vartheta_3}$&$100\%$\\ \hline
	    		\multirow{1}{*}{$\hbox{Gamma}(1,0.01)$}      
	    		&$\varphi$                 &$90\%$\\
	    		\multirow{1}{*}{$\hbox{Gamma}(1,0.001)$}      
	    		&$\phi$                 &$92\%$\\\hline\hline
	    		\multirow{6}{*}{$\hbox{Inverse Gamma}(0.01,0.01)$}               &$\omega^2$               &$92\% $\\
	    		&$\sigma^{2}_{\theta_1}$   &$100\%$\\
	    		&$\sigma^{2}_{\theta_2}$   &$100\%$\\
	    		&$\sigma^{2}_{\theta_3}$   &$100\%$\\
	    		&$\sigma^{2}_{\vartheta_1}$&$100\%$\\
	    		&$\sigma^{2}_{\vartheta_2}$&$100\%$\\
	    		&$\sigma^{2}_{\vartheta_3}$&$100\%$\\\hline
	    		\multirow{1}{*}{$\hbox{Gamma}(1,0.01)$}      
	    		&$\varphi$                 &$88\%$\\
	    		\multirow{1}{*}{$\hbox{Gamma}(1,0.001)$}      
	    		&$\phi$                 &$90\%$\\ \hline                                                 
	    	\end{tabular}
	    	\label{sim6}
	    \end{table}

	    \begin{table}[t]
	    	\centering
	    	\caption{Probability of coverage (PC) for the parameters of the proposed model in (\ref{modelo_proposto}) taken over the 50 independently generated datasets by applying the Inverse Scaled Chi-Squared priors for the parameters $\omega^2$, $\sigma^{2}_{\theta_i}$, and $\sigma^{2}_{\vartheta_i}$, $i=1,2,3$, and fixing Gamma$(1,0.01)$ and Gamma$(1,0.001)$ priors for the parameters $\varphi$ and $\phi$, respectively.}
	    	\begin{tabular}{cc|c}
	    		\hline
	    		Prior Distribution & Variance Component & \multicolumn{1}{c}{PC} \\ \hline
	    		\multirow{6}{*}{$\hbox{Scaled Inverse Chi-Squared} (0.0001,0.0001)$}          &$\omega^2$               &$90\% $\\
	    		&$\sigma^{2}_{\theta_1}$   &$98\%$ \\
	    		&$\sigma^{2}_{\theta_2}$   &$100\%$\\
	    		&$\sigma^{2}_{\theta_3}$   &$96\%$ \\
	    		&$\sigma^{2}_{\vartheta_1}$&$56\%$ \\
	    		&$\sigma^{2}_{\vartheta_2}$&$100\%$\\
	    		&$\sigma^{2}_{\vartheta_3}$&$80\%$ \\ \hline
	    		\multirow{1}{*}{$\hbox{Gamma}(1,0.01)$}      
	    		&$\varphi$                 &$90\%$\\
	    		\multirow{1}{*}{$\hbox{Gamma}(1,0.001)$}      
	    		&$\phi$                 &$88\%$\\ \hline\hline		
	    		
	    		\multirow{6}{*}{$\hbox{Scaled Inverse Chi-Squared}(0.001,0.001)$}            &$\omega^2$               &$90\% $\\
	    		&$\sigma^{2}_{\theta_1}$   &$100\%$\\
	    		&$\sigma^{2}_{\theta_2}$   &$100\%$\\
	    		&$\sigma^{2}_{\theta_3}$   &$100\%$\\
	    		&$\sigma^{2}_{\vartheta_1}$&$78\%$ \\
	    		&$\sigma^{2}_{\vartheta_2}$&$100\%$\\
	    		&$\sigma^{2}_{\vartheta_3}$&$80\%$ \\ \hline
	    		\multirow{1}{*}{$\hbox{Gamma}(1,0.01)$}      
	    		&$\varphi$                 &$90\%$\\
	    		\multirow{1}{*}{$\hbox{Gamma}(1,0.001)$}      
	    		&$\phi$                 &$92\%$\\ \hline\hline		
	    		\multirow{6}{*}{$\hbox{Scaled Inverse Chi-Squared}(0.01,0.01)$}               &$\omega^2$               &$90\% $\\
	    		&$\sigma^{2}_{\theta_1}$   &$100\%$\\
	    		&$\sigma^{2}_{\theta_2}$   &$100\%$\\
	    		&$\sigma^{2}_{\theta_3}$   &$100\%$\\
	    		&$\sigma^{2}_{\vartheta_1}$&$96\%$ \\
	    		&$\sigma^{2}_{\vartheta_2}$&$100\%$\\
	    		&$\sigma^{2}_{\vartheta_3}$&$96\%$ \\ \hline
	    		\multirow{1}{*}{$\hbox{Gamma}(1,0.01)$}      
	    		&$\varphi$                 &$88\%$\\
	    		\multirow{1}{*}{$\hbox{Gamma}(1,0.001)$}      
	    		&$\phi$                 &$90\%$\\ \hline		                                                  
	    	\end{tabular}
	    	\label{sim7}
	    \end{table}

\FloatBarrier

\subsection{Sensitivity Analysis of the Number of Locations and Sample Size of Repeated Measurements}\label{sim2}
		
In this section, we conducted a simulation study to assess the impact of the number of locations ($n$) and the number of repeated measurements per block ($J_i$, $i=1,2,3$) on the parameter estimates. This analysis holds relevance to the real dataset discussed in Section \ref{dados_reais}, where $n$ represents the count of meteorological stations evaluated, and $J_i$, $i=1,2,3$, denotes the number of years assessed within each climatic block.

For these simulation studies, we specified a set of prior distributions for the parameters, as detailed in Table \ref{sim4_hiper}. Across all scenarios, we maintained a burn-in period of $120,000$ iterations with a sampling interval of $100$ iterations. These simulation studies provide insights into how the choices of $n$ and $J_i$ affect the parameter estimates of the model.

\begin{table}[htb]
\centering
\caption{Prior Distributions Used in the Simulation Study to Evaluate the Effect of Changing the Number of Locations ($n$) and the Number of Repeated Measurements per Block ($J_i$, $i=1,2,3$) on Parameter Estimates of the Proposed Model in \eqref{modelo_proposto}.}
\begin{tabular}{cc}
	\hline 
				Parameters & Prior Distribution\\
				\hline 
				$\mbox{\boldmath $\beta$}$& $\mathbf{N}(\mathbf{0},10^3 \mathbf{I})$\\
				$\omega^2$&$\hbox{Inverse Gamma}(0.01,0.01)$\\
				$\sigma^{2}_{\theta_i}$&$\hbox{Inverse Gamma}(0.01,0.01)$\\
				$\sigma^{2}_{\vartheta_i}$ & $\hbox{Inverse Gamma}(0.01,0.01)$\\
				$\varphi$ & $\hbox{Gamma}(1,0.01)$\\
				$\phi$ & $\hbox{Gamma}(1,0.001)$\\
				\hline 
			\end{tabular}
			\label{sim4_hiper}
		\end{table}
		
In the simulation study discussed in Section \ref{sim1}, we investigated the impact of both the number of locations ($n$) and the number of time periods ($\tau$) on the parameter estimates. We generated 50 distinct datasets, each comprising 112 locations assessed over 9 time periods. These geographical configurations were designed to mimic real data and were augmented with random values within specified ranges of variation. For the repeated measurements, we utilized $J_1=12$ and $J_2=J_3=9$, while the model parameters were selected in accordance with Table \ref{reais}, incorporating $\kappa=0.2$, $K_{\mu}=9$, $K_{\zeta}=5$, and $K_{\Gamma}=7$. It is noteworthy that we maintained the condition $J_2=J_3$ in all simulations, consistent with observations in the actual dataset.

\begin{figure}[t]
\centerline{\includegraphics[width=510pt,height=18pc]{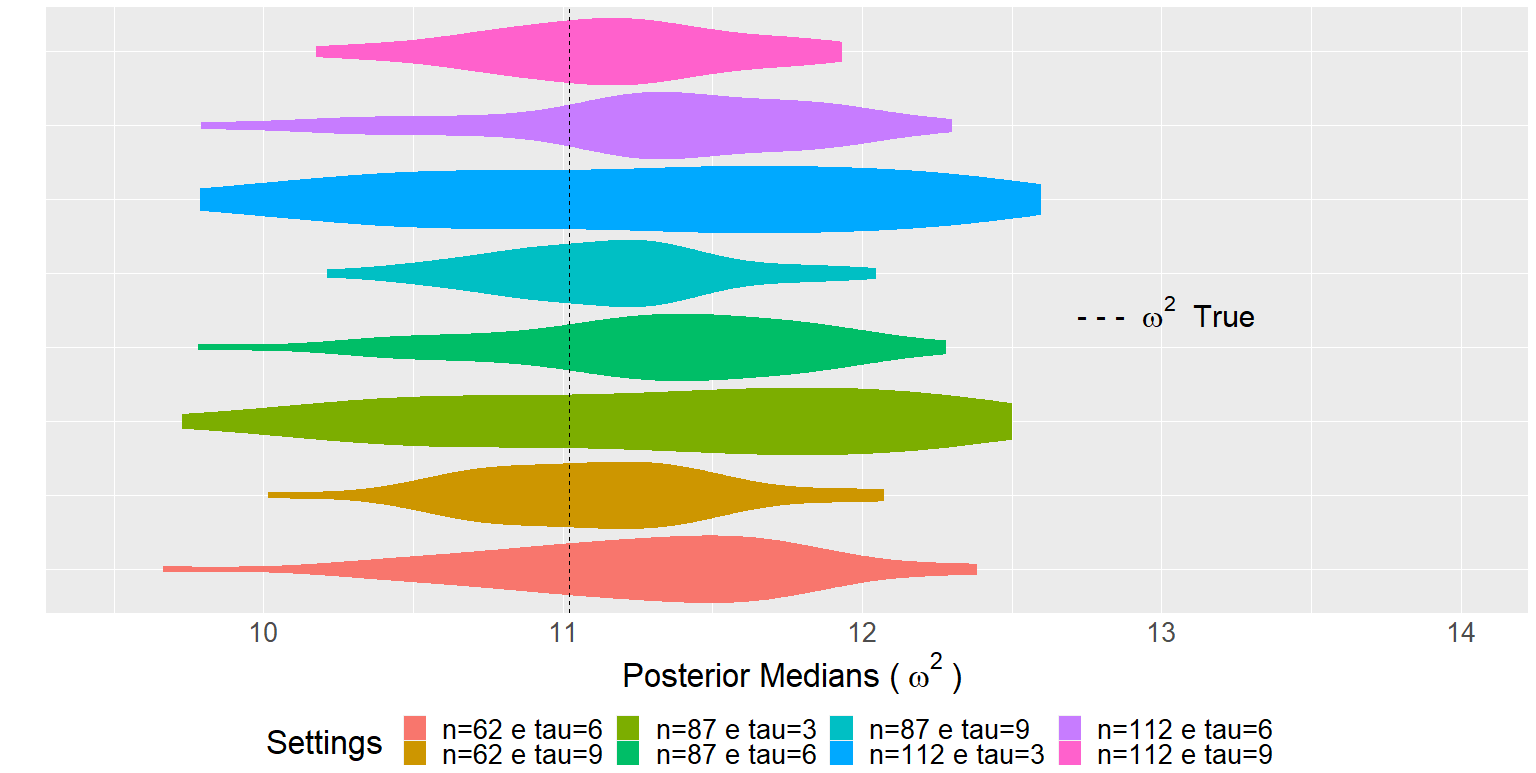}}
			\caption{Posterior medians for the parameters $\omega^2$, $\varphi$, and $\phi$, taken across the 50 independently simulated datasets while varying the number of locations ($n$) between 62, 87, and 112, and the number of observed periods ($\tau$) between 3, 6, and 9.\label{fig_omega2_median_1}}
\end{figure}

Figure \ref{fig_omega2_median_1} displays the posterior median of the parameter $\omega^2$ under different scenarios, where we vary the number of locations ($n$) between 62, 87, and 112, and the number of observed months ($\tau$) among 3, 6, and 9. The configuration with $n=62$ and $\tau=3$ is discarded due to insufficient chain convergence. Across all scenarios, the posterior estimates are consistently concentrated around the true values of the parameters. Increasing both $n$ and $\tau$ results in more precise and less variable estimates, indicating that increasing either of these dimensions enhances the quality of the estimates. In summary, increasing $n$ and/or $\tau$ leads to more precise and stable estimates of the evaluated parameters.

\begin{figure}[t]
\centerline{\includegraphics[width=510pt,height=18pc]{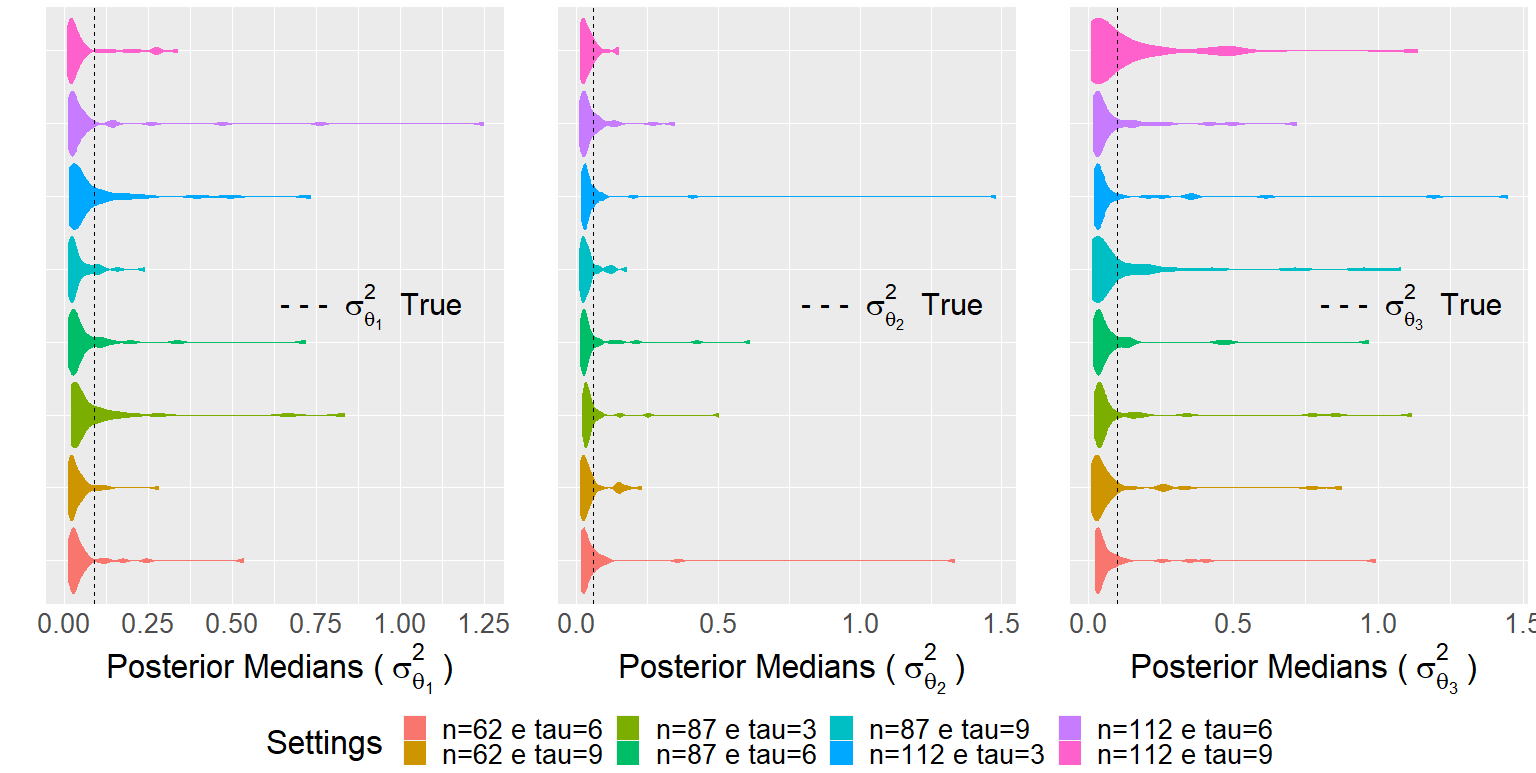}}
		\caption{Posterior medians for the parameters $\sigma^{2}_{\theta_i}$, $i=1,2,3$, taken across the 50 independently simulated datasets while varying the number of locations ($n$) between 62, 87, and 112, and the number of observed periods ($\tau$) between 3, 6, and 9.\label{fig_sigma2_theta_median_1}}
\end{figure}

\begin{figure}[t]
\centerline{\includegraphics[width=510pt,height=18pc]{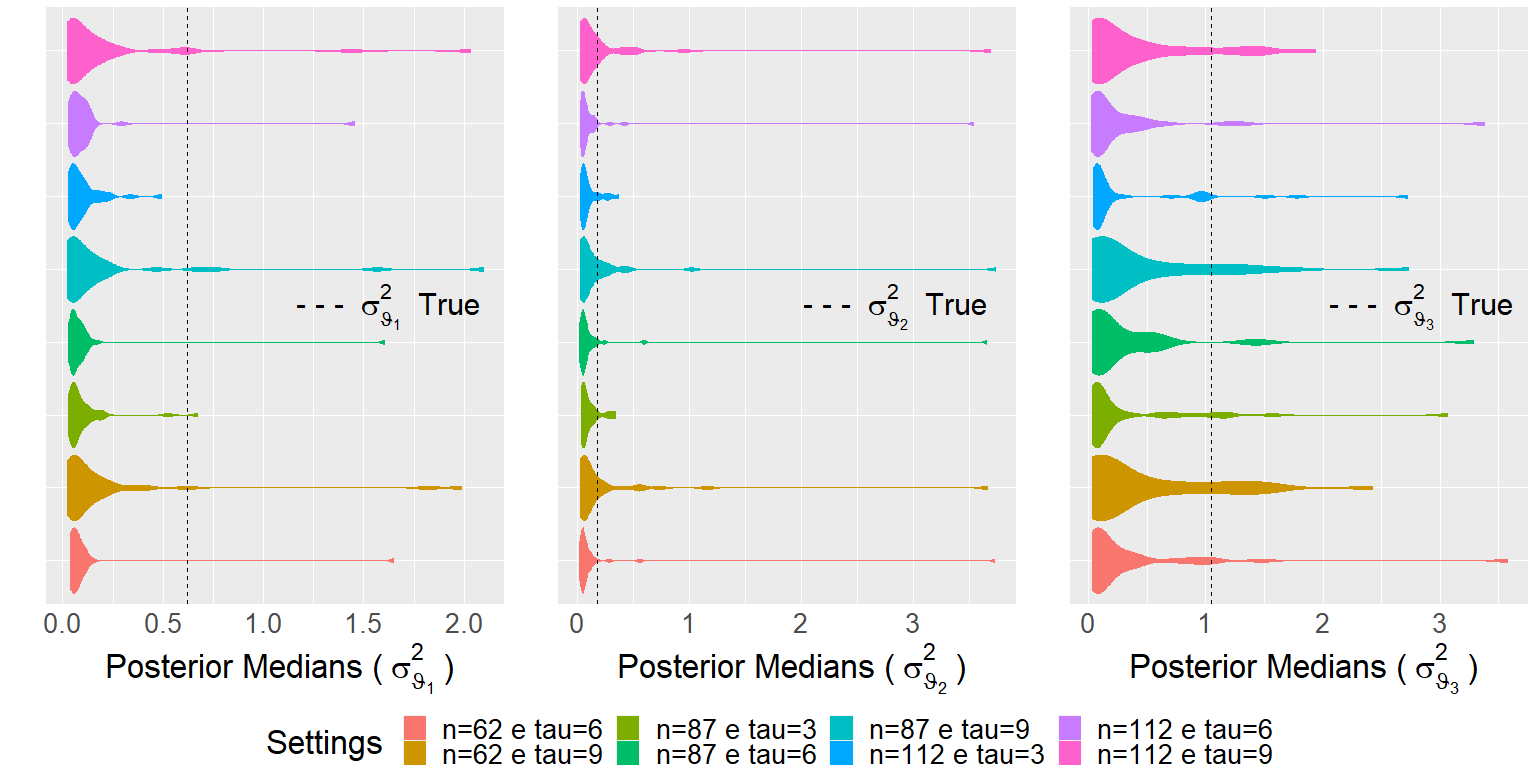}}
		\caption{Posterior medians for the parameters $\sigma^{2}_{\vartheta_i}$, $i=1,2,3$, taken across the 50 independently simulated datasets while varying the number of locations ($n$) between 62, 87, and 112, and the number of observed periods ($\tau$) between 3, 6, and 9.\label{fig_sigma2_vartheta_median_1}}
\end{figure}

The estimates of the variance components for spatial and temporal effects, illustrated in Figures \ref{fig_sigma2_theta_median_1} and \ref{fig_sigma2_vartheta_median_1}, respectively, represented by the posterior medians of the parameters $\sigma^{2}_{\theta_i}$ and $\sigma^{2}_{\vartheta_i}$, exhibit asymmetric behavior, often leaning toward underestimating the true parameter values. Moreover, it is worth noting that increasing either the number of locations ($n$) or the number of repeated measurements per block ($J_i$, where $i=1,2,3$) does not seem to result in a significant improvement in the precision of these estimates. In summary, the dimensions $n$ and $J_i$ evaluated in this study do not directly contribute to enhancing the precision of the estimates for variance components related to spatial and temporal effects. In other words, a much larger increase in these dimensions may be necessary to observe a significant improvement in the precision of these estimates.

Table \ref{sim_parte1} presents the coverage probabilities for the parameters in the proposed model \eqref{modelo_proposto}, obtained from this simulation study. The results consistently show that the coverage probabilities meet acceptable standards,  consistently exceeding $90\%$, thus confirming the quality of the estimation process.

\begin{table}[httb]
			\centering
			\caption{The table displays the coverage probabilities for the parameters of the proposed model in (\ref{modelo_proposto}), based on 50 independently generated datasets, with variations in the number of locations ($n$) between 62, 87, and 112, and the number of observed months ($\tau$) between 3, 6, and 9.}
			\footnotesize\resizebox{16.1cm}{!}{%
				\begin{tabular}{c|ccc|ccc|ccc}
					\hline
					\multicolumn{1}{c}{\multirow{1}{*}{}} & \multicolumn{9}{c}{Coverage probabilities} \\ \cline{2-10}
					\multicolumn{1}{c}{\multirow{2}{*}{}} & \multicolumn{3}{c|}{$\tau=3$}  & \multicolumn{3}{c|}{$\tau=6$} & \multicolumn{3}{c}{$\tau=9$} \\  
					\multicolumn{1}{c}{Parameters} & $n=62$ & $n=87$ & $n=112$ & $n=62$ & $n=87$ & $n=112$ & $n=62$ & $n=87$ & $n=112$ \\ \hline
					\multirow{9}{*}{}         
					$\omega^2$               &$--$  &$100\%$    &$98\%$ &$96\%$  &$98\%$  &$96\%$&$96\%$ &$100\%$&$100\%$\\  
					$\sigma^{2}_{\theta_1}$   &$--$  &$100\%$    &$100\%$&$100\%$ &$100\%$ &$100\%$&$100\%$ &$100\%$&$100\%$\\ 
					$\sigma^{2}_{\theta_2}$   &$--$  &$100\%$    &$100\%$&$100\%$ &$100\%$ &$100\%$&$100\%$ &$100\%$ &$100\%$\\   
					$\sigma^{2}_{\theta_3}$   &$--$  &$100\%$    &$100\%$&$100\%$ &$100\%$ &$100\%$&$100\%$ &$100\%$ &$100\%$\\   
					$\sigma^{2}_{\vartheta_1}$&$--$  &$100\%$    &$100\%$&$100\%$ &$100\%$ &$100\%$&$100\%$ &$100\%$ &$100\%$\\
					$\sigma^{2}_{\vartheta_2}$&$--$  &$100\%$    &$100\%$&$100\%$ &$100\%$ &$100\%$&$100\%$ &$100\%$ &$100\%$\\
					$\sigma^{2}_{\vartheta_3}$&$--$  &$100\%$    &$100\%$&$100\%$ &$96\%$  &$96\%$&$100\%$ &$98\%$ &$98\%$\\  
					$\varphi$                &$--$  &$92\%$     &$92\%$ &$96\%$  &$94\%$  &$90\%$&$94\%$  &$92\%$ &$90\%$\\   
					$\phi$                    &$--$  &$100\%$     &$100\%$&$96\%$  &$96\%$  &$96\%$&$98\%$ &$98\%$ &$100\%$\\
					\hline
			\end{tabular}}
			\label{sim_parte1}
		\end{table}
		
In the second simulation study, we explored the impact of varying the number of repeated measurements ($J_1$, $J_2$, and $J_3$) within each of the three blocks on the parameter estimates in the proposed model. We generated 50 independent datasets with $J_1=12$, $J_2=J_3=9$, $n=87$, and $\tau=6$. The geographical configurations were based on real data, and the parameter values followed those in Table \ref{reais}. Specifically, the configuration included $\kappa=0.2$, $K_{\mu}=9$, $K_{\zeta}=5$, and $K_{\Gamma}=7$. This study allowed us to evaluate how increasing the number of repeated measurements impacts the parameter estimates of the model.

\begin{figure}[t]
\centerline{\includegraphics[width=510pt,height=18pc]{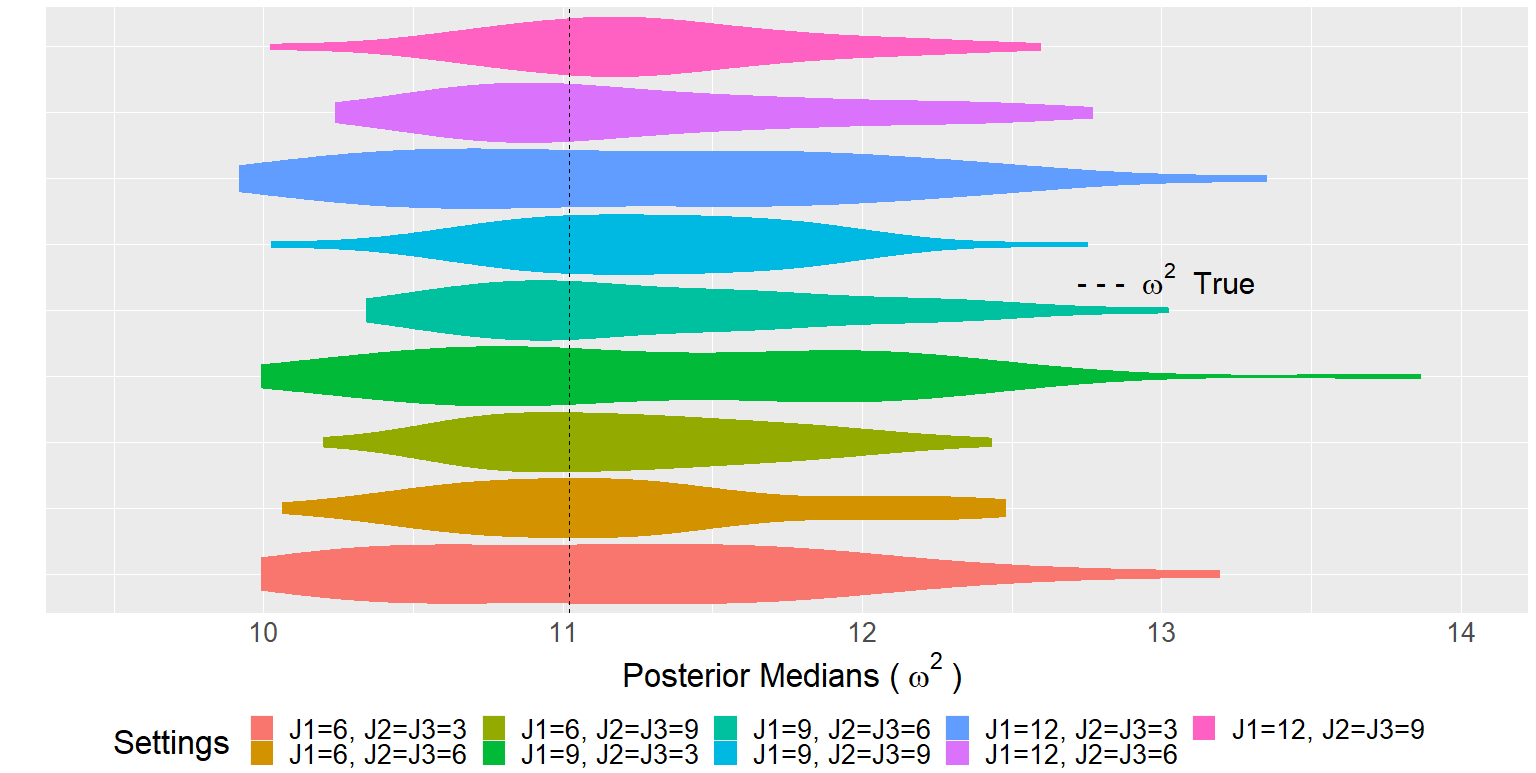}}
			\caption{Posterior medians for the parameters $\omega^2$, $\varphi$, and $\phi$, taken across the 50 independently simulated datasets while varying the number of repeated measurements per block, $J_i$, where $i=1,2,3$. \label{fig_omega2_median_2}}
\end{figure}

Figure \ref{fig_omega2_median_2} illustrates the posterior medians of the parameters $\omega^2$, $\varphi$, and $\phi$ in the model \eqref{modelo_proposto}, derived from 50 simulated datasets. The study varied the number of repeated measurements in each block of the model, either increasing or decreasing it by $3$ units relative to the dimension of the real data ($J_1=9$, $J_2=J_3=6$).

In general, when we maintain the number of measurements constant in one of the model's dimensions (i.e., $J_1$ or $J_2=J_3$) while increasing the other dimension, the posterior medians tend to cluster closer to the true parameter values. This implies that augmenting the number of measurements in a specific dimension enhances the precision of parameter estimates.

\begin{figure}[t]
\centerline{\includegraphics[width=510pt,height=18pc]{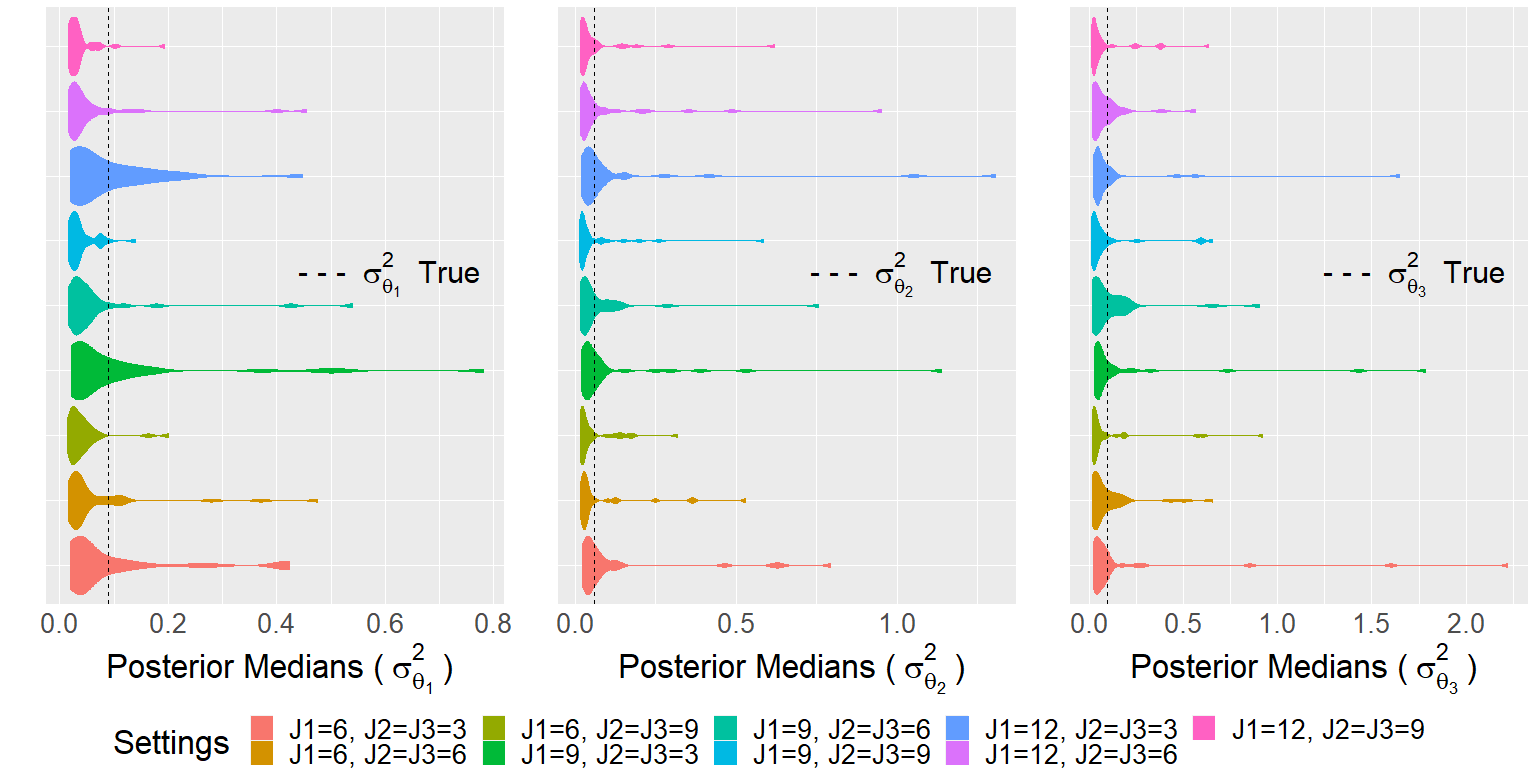}}
			\caption{Posterior medians for the parameters $\sigma^{2}_{\theta_i}$, $i=1,2,3$, taken across the 50 independently simulated datasets while varying the number of repeated measurements per block, $J_i$, where $i=1,2,3$. \label{fig_sigma2_theta_median_2}}
\end{figure}

\begin{figure}[t]
\centerline{\includegraphics[width=510pt,height=18pc]{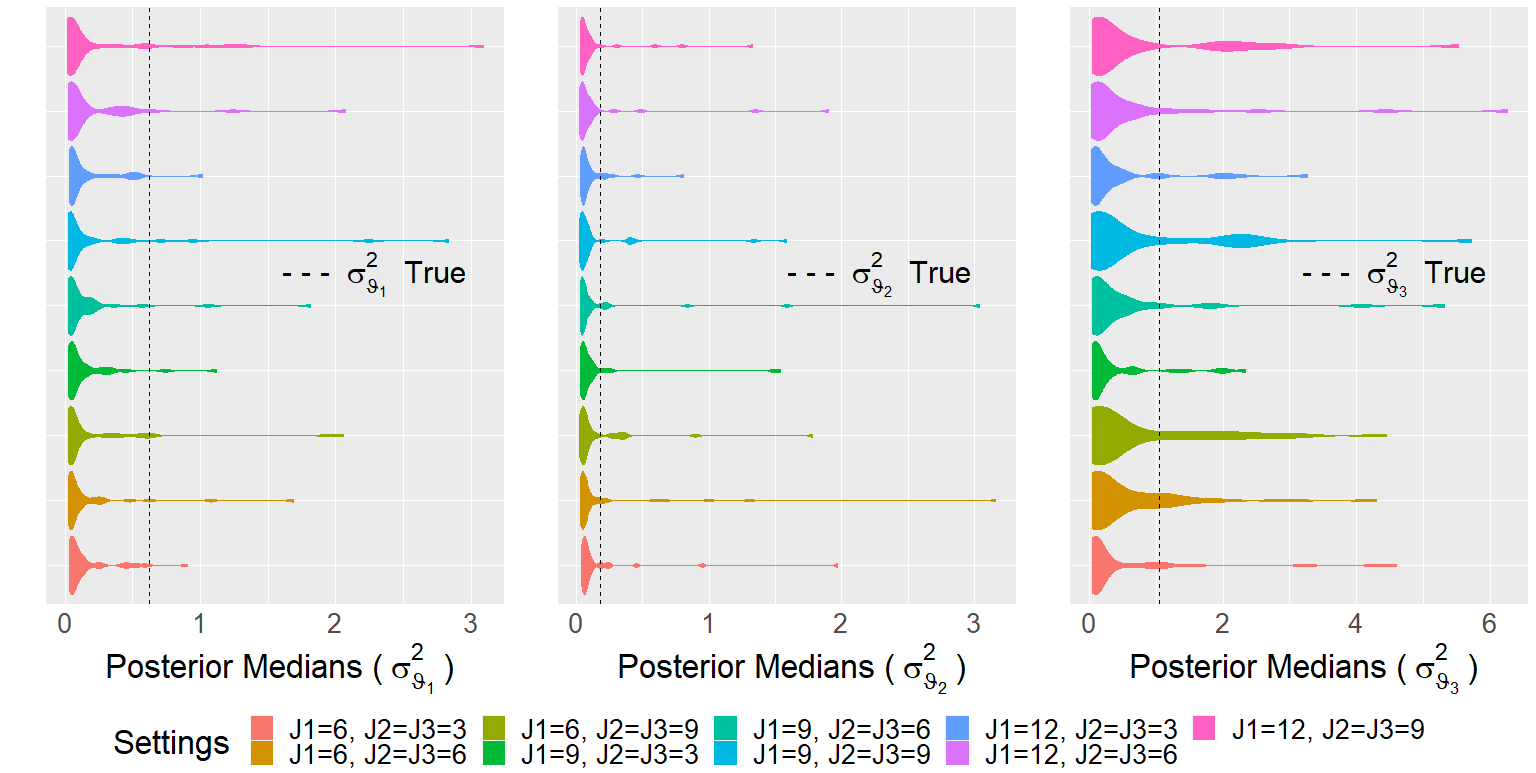}}
			\caption{Posterior medians for the parameters $\sigma^{2}_{\vartheta_i}$, $i=1,2,3$, taken across the 50 independently simulated datasets while varying the number of repeated measurements per block, $J_i$, where $i=1,2,3$. \label{fig_sigma2_vartheta_median_2}}
\end{figure}

The posterior medians for the parameters $\sigma^{2}_{\theta_i}$ and $\sigma^{2}_{\vartheta_i}$, which correspond to the variance components of the spatial and temporal effects, respectively, did not show a notable improvement in estimation accuracy with an increase in the number of repeated measurements, as depicted in Figures \ref{fig_sigma2_theta_median_2} and \ref{fig_sigma2_vartheta_median_2}. This suggests that increasing the number of repeated measurements does not have a significant influence on the precision of estimates for these variance components. However, this influence may become significant with a substantial increase in the number of repeated measurements.

\FloatBarrier
 \subsection{Selection of the Parameter \texorpdfstring{$\kappa$}{TEXT}}\label{sim3}

We conducted a simulation study to assess the ability of the model selection criteria LPML and DIC7 to accurately recover the true value of the parameter $\kappa$ as defined in \eqref{matern}, based on the goodness of fit of the model.

For this purpose, we generated 100 independent datasets for each value of $\kappa \in U=\{0.2,0.5,1,$ $1.5,2\}$, using the parameter values specified in Table \ref{reais}. These parameters were kept constant at $K_{\mu}=9$, $K_{\zeta}=5$, $K_{\Gamma}=7$, $\tau=6$, $J_1=9$, and $J_2=J_3=6$. The prior distributions used in this study are detailed in Table \ref{sim4_hiper}.

For each generated dataset, we fitted the proposed model in \eqref{modelo_proposto} with $\kappa \in U$ fixed. Subsequently, we used the LPML and DIC7 criteria to determine the best fit. Ideally, the value of $\kappa$ indicated by the model selection criteria should match the value used to generate the respective dataset. The results are presented in Table \ref{sim5}. It can be observed that both criteria exhibited accuracy rates exceeding $98\%$, indicating that the LPML and DIC7 criteria are effective in estimating the parameter $\kappa$.

\begin{table}[h]
\centering
\caption{Percentage of correct selections by the model selection criteria LPML and DIC7 associated with the choice of the true value of $\kappa$ described in the Matérn correlation structure. (\ref{matern}).}
			\begin{tabular}{ccc}
				\hline
				\multirow{2}{*}{True value of \texorpdfstring{$\kappa$}{TEXT} } & \multicolumn{2}{c}{Accuracy ($\%$)} \\ & LPML & DIC7 \\ \hline
				0.2 & 100 $\%$ & 100 $\%$ \\ 
				0.5 & 99 $\%$ & 99 $\%$ \\ 
				1.0 & 98 $\%$ & 98 $\%$ \\ 
				1.5 & 98 $\%$ & 98 $\%$ \\
				2.0 & 99 $\%$ & 99 $\%$ \\\hline
			\end{tabular}
			\label{sim5}
		\end{table}

\subsection{Predictive Capability of the Spatial-Temporal Functional Model with Block Structure and Repeated Measures} \label{sim_4}

We considered the predictive distribution to account for uncertainty when making predictions for an experimental unit that was not included in the original model fitting. In broad terms, we use the posterior distribution to evaluate the uncertainty associated with a future observation by integrating the likelihood over the posterior distribution of parameters. Thus, for an unobserved response $\mathbf{y}_{new}$, the posterior predictive distribution, based on the model proposed in \eqref{modelo_proposto}, can be expressed as
		
\begin{eqnarray}\label{dist_preditiva}
			p(\mathbf{y}_{new}|\mathbf{y})&= \int_{\mbox{\boldmath $\xi$}}\int_{\mbox{\boldmath $\vartheta$}}\int_{\mathbf{\Theta}}p(\mathbf{y}_{new}|\mathbf{y},\mbox{\boldmath $\xi$},\mbox{\boldmath $\vartheta$},\mathbf{\Theta})p(\mbox{\boldmath $\xi$},\mbox{\boldmath $\vartheta$},\mathbf{\Theta}|\mathbf{y})d\mbox{\boldmath $\xi$}d\mbox{\boldmath $\vartheta$}d\mathbf{\Theta}\nonumber\\
			&=\int_{\mbox{\boldmath $\xi$}}\int_{\mbox{\boldmath $\vartheta$}}\int_{\mathbf{\Theta}}p(\mathbf{y}_{new}|\mbox{\boldmath $\xi$},\mbox{\boldmath $\vartheta$},\mathbf{\Theta})p(\mbox{\boldmath $\xi$},\mbox{\boldmath $\vartheta$},\mathbf{\Theta}|\mathbf{y})d\mbox{\boldmath $\xi$}d\mbox{\boldmath $\vartheta$}d\mathbf{\Theta},
		\end{eqnarray}
		
		\noindent in which the second equality considers the conditional independence between $\mathbf{y}_{new}$ and $\mathbf{y}$ and, therefore, $p(\mathbf{y}_{new}|\mathbf{y})=E_{\mbox{\boldmath $\xi$},\mbox{\boldmath $\vartheta$},\mathbf{\Theta}}(p(\mathbf{y}_{new}|\mbox{\boldmath $\xi$},\mbox{\boldmath $\vartheta$},\mathbf{\Theta}))$. Through the use of MCMC via Gibbs sampling, it is possible to obtain samples from the joint posterior distribution, as described in \cite{Gamerman2006}, by estimating the predictive distribution given by
  
		\begin{eqnarray}\label{est_dist_preditiva}
			\widehat{p(\mathbf{y}_{new}|\mathbf{y})}= \dfrac{1}{q} \displaystyle\sum_{h=1}^{q} p(\mathbf{y}_{new}|\mbox{\boldmath $\xi$}^{h},\mbox{\boldmath $\vartheta$}^{h},\mathbf{\Theta}^{h}),
		\end{eqnarray}
		
\noindent where $\mbox{\boldmath $\xi$}^{h}$, $\mbox{\boldmath $\vartheta$}^{h}$, and $\mathbf{\Theta}^{h}$, $h=1,...,q$, are samples from the posterior distribution $p(\mbox{\boldmath $\xi$},\mbox{\boldmath $\vartheta$},\mathbf{\Theta}|\mathbf{y})$ obtained through the Gibbs sampler.
		
To evaluate the predictive performance of the model proposed in (\ref{modelo_proposto}), we utilized parameters $\kappa=0.2$, $K_{\mu}=9$, $K_{\zeta}=5$, and $K_{\Gamma}=7$, along with the specified prior distributions from Table \ref{sim4_hiper}. Initially, we generated a new dataset with 113 locations ($n=113$) and $\tau=6$, including 13 repeated measurements for Block $i=1$ ($J_1=13$) and 10 repeated measurements for each of Blocks $i=2,3$ ($J_2=J_3=10$), mimicking the structure assessed in Subsection \ref{sim2}. To create a test data at a new location, we removed the last repeated measurement from each block in the database and then randomly omitted one of the original 113 locations, resulting in a training dataset with dimensions of $n=112$, $\tau=6$, $J_1=12$, and $J_2=J_3=9$.

\begin{figure}[h]
\centerline{\includegraphics[width=510pt,height=18pc]{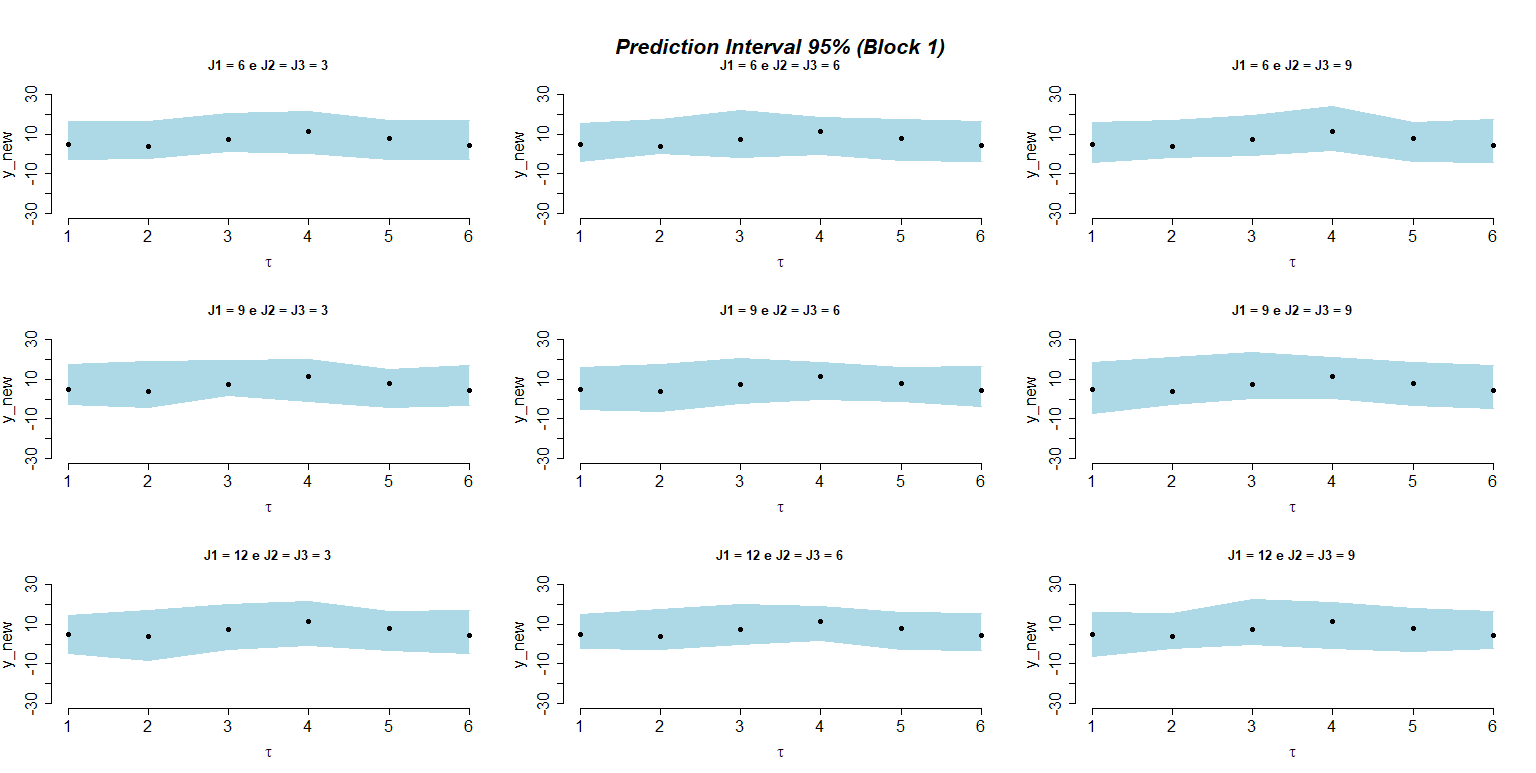}}
			\caption{$95\%$ prediction intervals for a test location evaluated in Block $i=1$, taken from the estimates of the proposed model in (\ref{modelo_proposto}), while varying $J_i$, $i=1,2,3$, using $\tau=6$, $\kappa=0.2$, $K_{\mu}=9$, $K_{\zeta}=5$, and $K_{\Gamma}=7$.}
			\vspace{-0.4cm}
			\label{pred_comp_v1_1}
\end{figure}

Figure \ref{pred_comp_v1_1} illustrates the $95\%$ prediction intervals evaluated at the test location in Block 1, obtained by varying the number of repeated measurements across all blocks, $J_i$, $i=1,2,3$, during the model training process. As observed, the test measurements are entirely contained within their respective prediction intervals. However, variations in the number of repeated measurements ($J_i$, $i=1,2,3$) within the blocks did not lead to a significant reduction in the width of the prediction intervals. Similar results were found for Blocks 2 and 3.	

In the subsequent study, we generated a dataset featuring 113 locations, $\tau=9$, and 10 repeated measurements for Block $i=1$, while Blocks $i=2$ and $i=3$ had 7 repeated measurements each, replicating the structure of the real data. To assess the predictive performance of the model, we randomly selected a test location and removed the last repeated measurement in each block, reducing the training set to a maximum of 112 locations. This resulted in $\tau=9$ with 9 repeated measurements for Block $i=1$ and 6 repeated measurements for Blocks $i=2$ and $i=3$. The primary objective here is to evaluate the model's predictive performance while randomly varying the values of $n$ and $\tau$ during the training process.

\begin{figure}[t]
\centerline{\includegraphics[width=510pt,height=18pc]{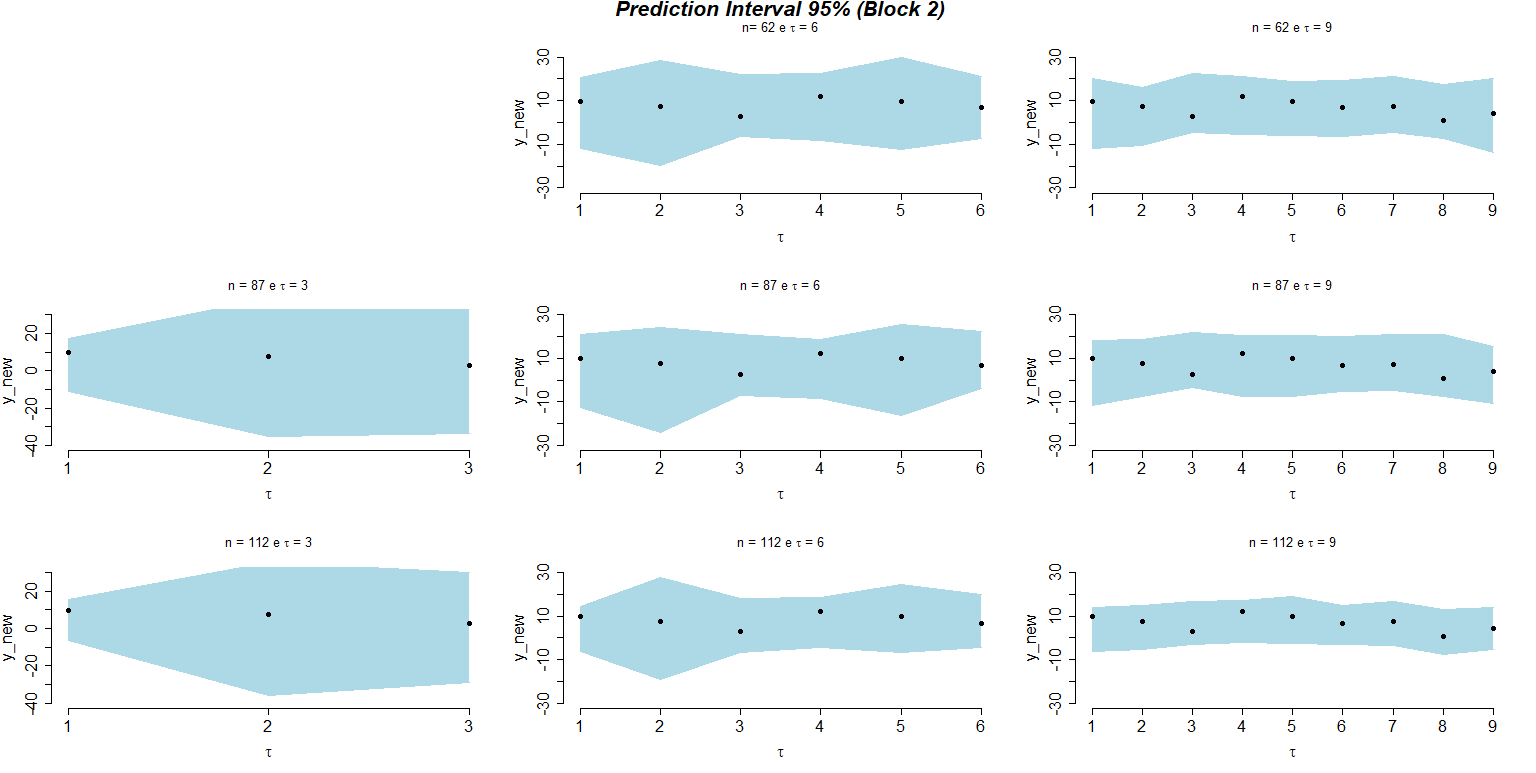}}
\caption{$95\%$ prediction intervals for a test location evaluated in Block $i=2$, obtained from the estimates of the proposed model in (\ref{modelo_proposto}), while varying $n$ and $\tau$, using $J_1=9$, $J_2=J_3=6$, $\kappa=0.2$, $K_{\mu}=9$, $K_{\zeta}=5$, and $K_{\Gamma}=7$.}
			\vspace{-0.4cm}
			\label{pred_comp_v2_2}
\end{figure}

Figure \ref{pred_comp_v2_2} presents $95\%$ prediction intervals for the scenarios in Block 2. It is worth noting that the scenario with $n=62$ and $\tau=3$ is not represented due to convergence issues. In all the depicted cases, the new measurements fall entirely within the prediction intervals. Moreover, increasing either $n$ or $\tau$ results in narrower prediction intervals, enhancing the predictive capability of the model. It is worthy mentioning that other similar simulations yielded consistent results with those presented here.

\FloatBarrier
\section{Applications of Precipitation Data: A Study of the State of Goiás, Brazil} \label{dados_reais}

\begin{figure}[h]
\centerline{\includegraphics[width=450pt,height=18pc]{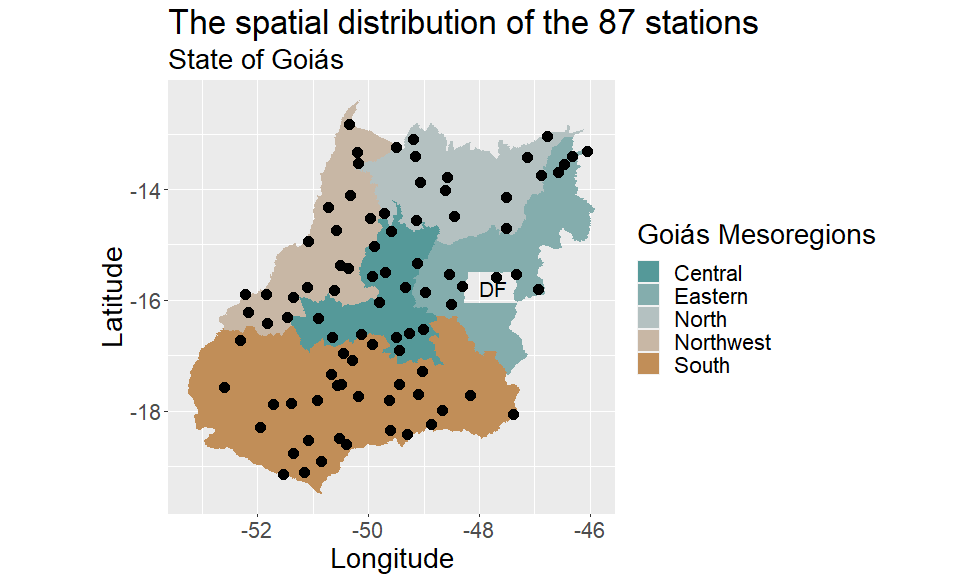}}
			\caption{Spatial distribution of the 87 meteorological stations under study in the state of Goiás.\label{mapa87}}
\end{figure}

In this investigation, the weather stations are situated within the state of Goiás, Brazil, which spans an area of 340,086 square kilometers. The region predominantly experiences a tropical savannah climate, characterized by two distinct seasons: the wet season and the dry season. The dataset comprises precipitation data collected from 87 stations distributed across five mesoregions within the area. These observations were recorded during the months from October to March and were sourced from various institutions, including INMET, ANA, SIMEHGO, and EMBRAPA. The geographical distribution of these stations is visually depicted in Figure \ref{mapa87}.

Meteorological data spanning the years 1980 to 2001, encompassing a total of 21 years, were categorized based on the climatic conditions associated with the El Niño-Southern Oscillation (ENSO) phenomenon. ENSO is characterized by fluctuations in sea surface temperatures (SST) in the Equatorial Pacific Ocean, as quantified by the Oceanic Niño Index (ONI) provided by the National Oceanic and Atmospheric Administration (NOAA). The classification takes into account ONI values, with values less than -0.5°C indicating predominantly La Niña years, values greater than 0.5°C signifying predominantly El Niño years, and values falling between -0.5°C and 0.5°C representing Neutral years (source: \citep{cliamte2019}).

Table \ref{classif_geral} presents the comprehensive classification of the dataset in relation to the predominant climatic effect. For a more detailed understanding, Table \ref{classif} describes the classification process for the period from 1980 to 1983. As we can observe, the classification composition for each period begins with the forecast for the months of August, September, and October (ASO) and ends with the forecast for the months of March, April, and May (MAM).

		\begin{table}[!http]
			\centering
			\caption{Classification of climatic effects for the period from 1980 to 2001.}
			\begin{tabular}{cccc}
				\hline 
				\multirow{2}{*}{} &
				\multicolumn{3}{c}{Predominant Climatic Effect}\\\cline{2-4}
				&La Niña & Neutral  & El Niño\\
				\hline 
				\multirow{9}{*}{Períodos}  & 1984-1985 & 1980-1991& 1982-1983\\
				& 1988-1989 & 1981-1982& 1986-1987\\
				& 1995-1996 & 1983-1984& 1987-1988\\
				& 1998-1999 & 1985-1986& 1991-1992\\
				& 1999-2000 & 1989-1990& 1994-1995\\
				& 2000-2001 & 1990-1991& 1997-1998\\
				& --- & 1992-1993& ---\\
				& --- & 1993-1994& ---\\
				& --- & 1996-1997& ---\\
				\hline 
			\end{tabular}
			\label{classif_geral}
		\end{table}

  \begin{table}[!http]
			\centering
			\caption{Illustration of the classification of predominant climatic effects for the period from 1980 to 1983.}
			\begin{tabular}{lllllllllll}
				\hline 
				\multirow{2}{*}{Period} &
				\multicolumn{10}{c}{Oceanic Niño Index (ONI)}\\\cline{2-11}
				&ASO & SON & OND & NDJ & DJF & JFM & FMA& MAM & Mean & Effect\\
				\hline 
				1980-1981 \vline&-0.1& 0.0    & 0.1  & 0.0    & -0.3 & -0.5 & -0.5 & -0.4 &-0.21 &Neutro\\
				1981-1982 \vline&-0.2& -0.1 & -0.2 & -0.1 & 0.0   & 0.1  & 0.2  & 0.5  &0.02& Neutro\\
				1982-1983 \vline&1.6 & 2.0    & 2.2  & 2.2  & 2.2  & 1.9  & 1.5  & 1.3  &1.86 &El Niño\\
				\hline 
			\end{tabular}
			\label{classif}
		\end{table}

\begin{figure}[ht]
\centerline{\includegraphics[width=510pt,height=18pc]{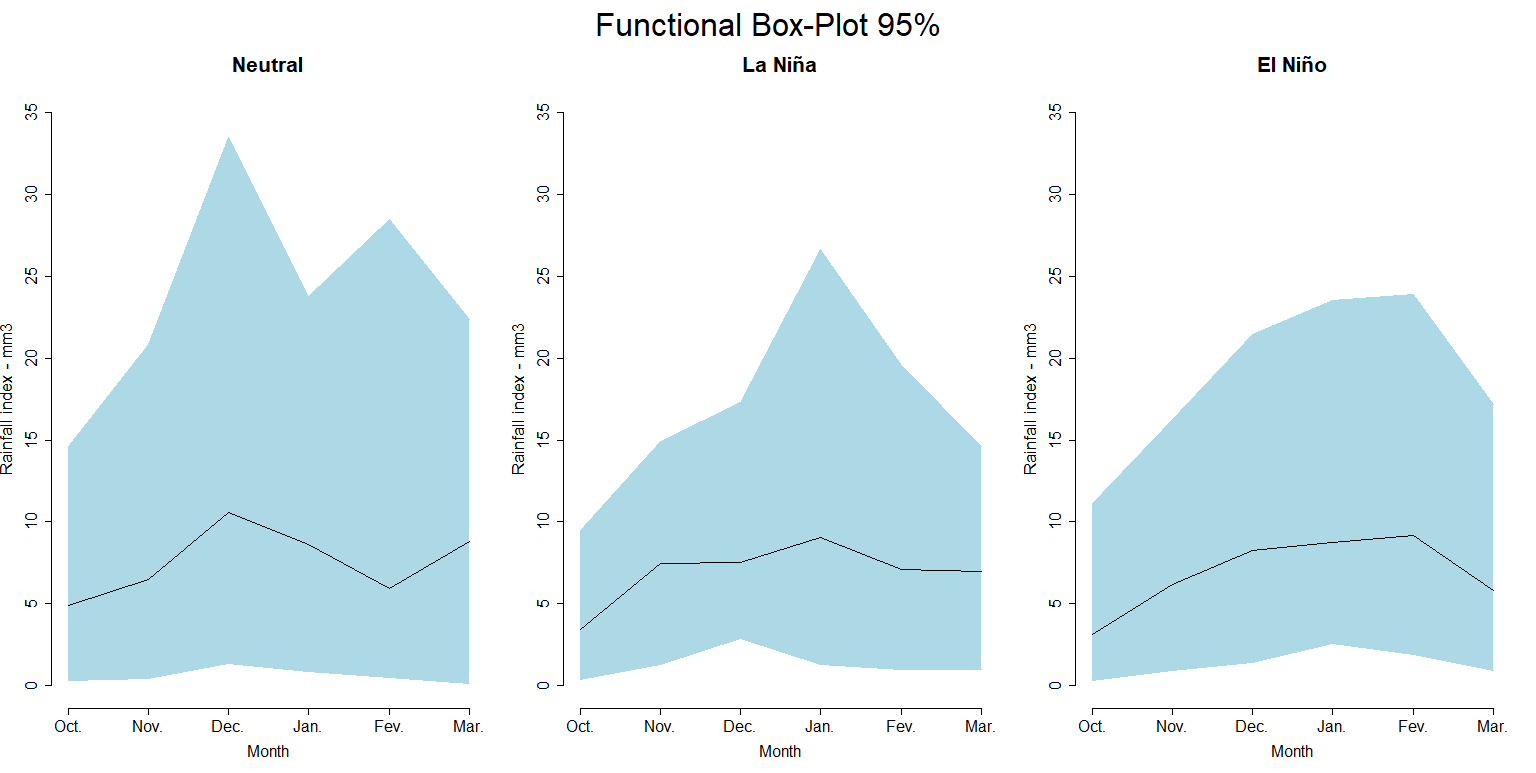}}
			\caption{Functional boxplot for the monthly average precipitation index ($mm^3$) with $95\%$ depth bands, evaluated across the climatic effect groups Neutral, La Niña, and El Niño, for the period from 1980 to 2001.\label{boxplot_periodos}}
\end{figure}

Figure \ref{boxplot_periodos} presents the functional boxplot, which illustrates the curves of monthly average precipitation across the different weather stations analyzed in this study, over the months and for different climatic conditions (Neutral, La Niña, and El Niño) from 1980 to 2001. The bold line represents the zero-level curve, indicating the median of the curves. Distinct seasonal patterns and different peaks of precipitation, depending on the climatic effect assessed, can be observed. This visual analysis suggests that the model described in (\ref{modelo_proposto}) is effective in capturing the variations in precipitation associated with the different climatic conditions in the dataset.

\FloatBarrier

\subsection{Fitting the Spatial-Temporal Functional Model with Block Structure and Repeated
Measures to Precipitation Data} \label{aplicacao}
     
Initially, we evaluated the parameter $\kappa$ within the Matérn correlation structure (\ref{matern}) across the range of values from 0.1 to 3.5. We set the upper limit at 3.5 based on previous simulation studies, which indicated that values beyond this threshold could compromise the accuracy of estimating $\kappa$ and impact the estimates of the parameter $\phi$, which is also part of this correlation structure. Additionally, we selected the numbers of bases for the expansions of the fixed functional ($K_{\mu}$), spatial ($K_{\zeta}$), and temporal ($K_{\Gamma}$) effects from the set ${5, 6, 7, 8, 9}$. Specifics regarding the prior distributions employed in this section are provided in Table \ref{sim4_hiper}.

Table \ref{sim4} summarizes the selected scenarios based on the LPML and DIC7 criteria. In all scenarios, the value of $\kappa$ was held constant at 0.2. However, the LPML and DIC7 criteria resulted in different configurations for the combinations of basis numbers used in the functional components of the model. Overall, the configuration $K_{\mu}=9$, $K_{\zeta}=5$, and $K_{\Gamma}=7$ stood out, ranking second and third in the LPML and DIC7 criteria, respectively. For this reason, this configuration were adopted in our study.

\begin{table}[htb]
    	\centering
\caption{LPML and DIC7 results obtained in the fitting of the proposed model in (\ref{modelo_proposto}) considering various values of $\kappa$ and numbers of bases used in the expansions of the fixed, spatial, and temporal functional effects when applied to the real dataset.}
    	\begin{tabular}{ccccc}
    		\hline 
    		Position & Criterion & Estimates & Number of Bases & $\kappa$ \\
    		\hline 
    		\multirow{2}{*}{1\textsuperscript{\underline{a}}}& LPML &-23940.0 &$K_{\mu}=9$, $K_{\zeta}=8$ e $K_{\Gamma}=9$ &0.2\\ &DIC7 &47925.4& $K_{\mu}=9$, $K_{\zeta}=5$ e $K_{\Gamma}=9$&0.2\\ \hline
    		\multirow{2}{*}{2\textsuperscript{\underline{a}}}& LPML & -23940.8 &$K_{\mu}=9$, $K_{\zeta}=5$ e $K_{\Gamma}=7$ &0.2\\ &DIC7 & 47925.5 & $K_{\mu}=9$, $K_{\zeta}=8$ e $K_{\Gamma}=7$&0.2\\ \hline
    		\multirow{2}{*}{3\textsuperscript{\underline{a}}}& LPML &-23941.3 &$K_{\mu}=9$, $K_{\zeta}=6$ e $K_l=9$ &0.2\\ &DIC7 & 47925.7& $K_{\mu}=9$, $K_{\zeta}=5$ e $K_{\Gamma}=7$&0.2\\ \hline
    		\multirow{2}{*}{4\textsuperscript{\underline{a}}}& LPML & -23942.6&$K_{\mu}=9$, $K_{\zeta}=5$ e $K_{\Gamma}=8$ &0.2\\ &DIC7 & 47925.9&$K_{\mu}=9$, $K_{\zeta}=8$ e $K_{\Gamma}=8$&0.2\\ \hline
    		\multirow{2}{*}{5\textsuperscript{\underline{a}}}& LPML &-23943.4 &$K_{\mu}=9$, $K_{\zeta}=5$ e $K_{\Gamma}=9$ &0.2\\ &DIC7 & 47926.4& $K_{\mu}=9$, $K_{\zeta}=9$ e $K_{\Gamma}=8$&0.2\\
    		\hline 
    	\end{tabular}
    	\label{sim4}
    \end{table}
    
We established prior distributions for the parameter vector $\xi$ of the model presented in (\ref{modelo_proposto}), as outlined in Table \ref{sim4_hiper}. The sampling procedure employed Gibbs sampling encompassing $1.5 \times 10^5$ steps, including a warm-up phase of $10^5$ steps and a sampling frequency of every $100$ steps. To ensure convergence and uncorrelated samples, we employed two chains for each parameter, initiating them from distinct starting points, as specified in Table \ref{chutes_iniciais}. Convergence assessment was carried out using the Gelman and Rubin statistic \citep{Gelman1992}, where values close to 1 indicate convergence. The results confirmed convergence for all estimated parameters, as detailed in Table \ref{criterio_conv}.

\begin{table}[htb]
   	\centering
   	\caption{Initial values used to generate two sets of Markov chains for the parameters of the model in (\ref{modelo_proposto}), when applied to precipitation data in the State of Goiás, used in the Gibbs sampler scheme.}
   	\begin{tabular}{lc|cc}
   		\hline
   		\multicolumn{2}{c}{\multirow{2}{*}{}}&\multicolumn{2}{c}{Initial Values} \\  \cline{3-4}
   		\multicolumn{2}{c}{\multirow{1}{*}{}}& Chain 1 & Chain 2 \\ \hline
   		\multirow{9}{*}{Parameters}
   		&$\omega^2$ & 0.1 & 100\\ 
   		&$\sigma^{2}_{\theta_1}$ & 0.01& 10\\ 
   		&$\sigma^{2}_{\theta_2}$&  0.01& 10 \\ 
   		&$\sigma^{2}_{\theta_3}$ &  0.01 & 10\\ 
   		&$\sigma^{2}_{\vartheta_1}$ & 0.01 & 10\\ 
   		&$\sigma^{2}_{\vartheta_2}$& 0.01 & 10\\ 
   		&$\sigma^{2}_{\vartheta_3}$ &  0.01 & 10\\ 
   		&$\varphi$ & 0.1 & 100\\ 
   		&$\phi$& 10 & 1000\\\hline
   	\end{tabular}
   	\label{chutes_iniciais}
   \end{table}
     
\begin{table}[htb]
    	\centering
    	\caption{The Gelman and Rubin statistic for the parameters of the model proposed in (\ref{modelo_proposto}) when applied to precipitation data in the state of Goiás.}
    \begin{tabular}{lc|c}
    \hline
    \multicolumn{2}{c}{\multirow{2}{*}{}} & \multicolumn{1}{c}{Gelman and Rubin statistic} \\  
    \hline
    \multirow{9}{*}{Parameters}
    &$\omega^2$                 & 1.007  \\ 
    &$\sigma^{2}_{\theta_1}$    & 1.000  \\ 
    &$\sigma^{2}_{\theta_2}$    & 1.002  \\ 
    &$\sigma^{2}_{\theta_3}$    & 1.029  \\ 
    &$\sigma^{2}_{\vartheta_1}$ & 1.004  \\ 
    &$\sigma^{2}_{\vartheta_2}$ & 1.012  \\ 
    &$\sigma^{2}_{\vartheta_3}$ & 0.999  \\ 
    &$\varphi$                  & 1.003  \\ 
    &$\phi$                     & 1.005  \\ \hline
    &$\mbox{\boldmath $\xi$}$   & 1.030  \\ \hline                         
\end{tabular}
\label{criterio_conv}
\end{table}

\begin{table}[h]
    	\centering
    	\caption{A posteriori estimates of the parameters composing the model proposed in (\ref{modelo_proposto}): median and standard deviation, obtained from samples of two Markov chains generated for each of the parameters of the proposed model.}
    	\begin{tabular}{cl|cc}
    		\hline
    		\multicolumn{2}{c}{\multirow{2}{*}{}} & \multicolumn{2}{c}{Posterior Statistics} \\\cline{3-4} 
    		\multicolumn{2}{c}{}                 & Median & Standard Deviation \\ \hline 
    		\multirow{9}{*}{Parameters}         
    		&$\omega^2$ 	&$10.989$ &$0.582$\\ 
    		&$\sigma^{2}_{\theta_1}$ &$ 0.006$ &$0.091$\\
    		&$\sigma^{2}_{\theta_2}$ &$0.007$  &$0.105$\\
    		&$\sigma^{2}_{\theta_3}$ &$0.006$  &$0.123$\\
    		&$\sigma^{2}_{\vartheta_1}$ &$0.035$  &$0.992$\\
    		&$\sigma^{2}_{\vartheta_2}$ &$0.008$  &$0.313$\\
    		&$\sigma^{2}_{\vartheta_3}$ &$ 0.037$ &$1.358$\\
    		&$\varphi$ &$2.030$  &$0.079$\\
    		&$\phi$ &$697.908$&$95.379$\\ \hline
      \end{tabular}
    	\label{est_post}
    \end{table}
    
\begin{figure}[ht]
\centerline{\includegraphics[width=450pt,height=18pc]{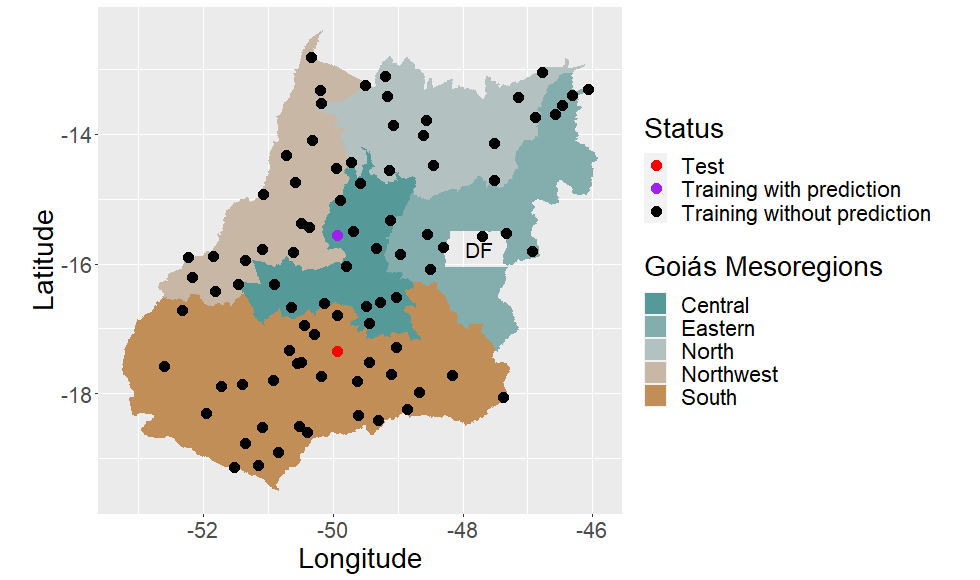}}
    	\caption{Spatial distribution of meteorological stations classified as ``Training'', ``Prediction Test'', and ``Test without prediction'', used to verify the results provided by the proposed model in (\ref{modelo_proposto}), when applied to precipitation data in the State of Goiás. \label{Mapa_87_media_mensal_preditas}}
\end{figure}

To build the forecasting scenarios, we randomly selected a station classified as ``Training with prediction'' located in the Central mesoregion of the state of Goiás and collected additional data from a meteorological station in the South mesoregion, referred to as ``Test.'' Figure \ref{Mapa_87_media_mensal_preditas} shows the geographical distribution of these stations, and Table \ref{estações_avaliadas} provides detailed information about their temporal and spatial characteristics.

   \begin{table}[!http]
    	\centering
    	\caption{Municipality, latitude, longitude, and classification as ``Test'' or ``Training with prediction'' of the meteorological stations used to verify the results provided by the proposed model in (\ref{modelo_proposto}), when applied to precipitation data in the State of Goiás.\label{estações_avaliadas}}
    	\begin{tabular}{cccccc}
    		\hline 
    		\multirow{2}{*}{Goiás Mesoregion} &
    		\multicolumn{4}{c}{Evaluated Stations}\\\cline{2-5}
    		& Municipality & Latitude & Longitude & State\\
    		\hline
    		\multirow{1}{*}{Central} & Itapuranga & -15.56 & -49.94 & Training with prediction \\
    		\hline
    		\multirow{1}{*}{South} & Edéia    & -17.34&-49.93  & Test \\
    		\hline
    	\end{tabular}
    \end{table}

\begin{figure}[ht]
\centerline{\includegraphics[width=450pt,height=18pc]{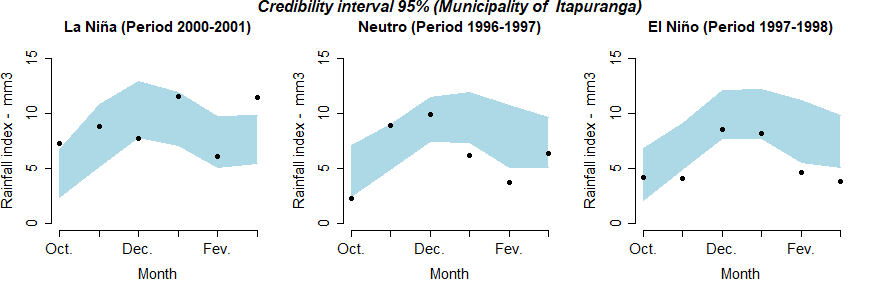}}
    	\caption{Credibility intervals for the estimated monthly average rainfall indices for the meteorological station ``Training with prediction'', located in the municipality of Itapuranga-GO, over the last observed periods for each of the climatic effects. \label{credibilidade_itapuranga}}
\end{figure}

Credibility intervals were established for the periods 1996-1997, 1997-1998, and 2000-2001, representing Neutral, La Niña, and El Niño climate events, respectively. Figure \ref{credibilidade_itapuranga} illustrates that, irrespective of climatic fluctuations, the observed data points consistently fall within the credibility intervals. This confirms the model's reliability and its ability to provide accurate results under various climatic conditions in this study.

\begin{figure}[ht]
\centerline{\includegraphics[width=510pt,height=18pc]{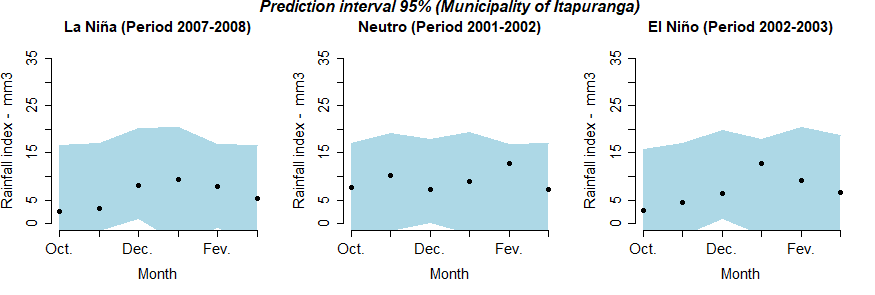}}
\caption{Prediction intervals for the observed monthly average rainfall indices for the meteorological station ``Training with prediction'', located in the municipality of Itapuranga-GO, over the climatic effects observed after the period of 2000-2001. \label{predicao_itapuranga}}
\end{figure}

According to the classification by \cite{cliamte2019}, the Neutral, El Niño, and La Niña climate events occurred after the period analyzed in this study, specifically in the years 2001-2002, 2002-2003, and 2007-2008. These periods were used to assess the predictive capability of the model presented in this study. Figure \ref{predicao_itapuranga} illustrates the prediction intervals for the station categorized as ``Training with prediction'', as outlined in Table \ref{estações_avaliadas}. The prediction intervals have a wide range; however, it is possible to observe that all observed average rainfall values are covered by the prediction interval.

\begin{figure}[ht]
\centerline{\includegraphics[width=510pt,height=18pc]{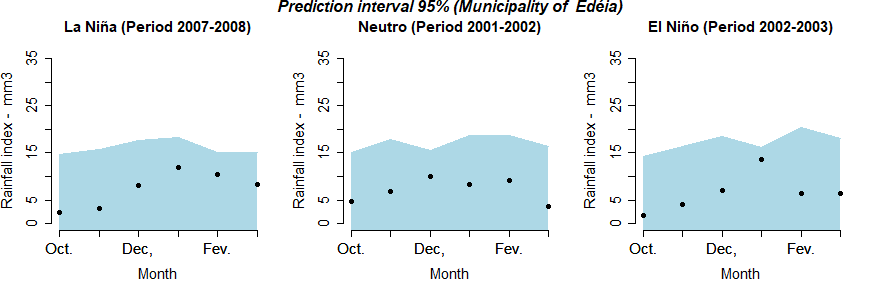}}
    	\caption{Prediction intervals for the observed monthly average rainfall indices for the meteorological station ``Test'' located in the municipality of Edéia-GO, over each of the climatic effects observed after the period of 2000-2001.\label{predicao_goias}}
\end{figure}

In Figure \ref{predicao_goias}, the adjusted prediction intervals for the station categorized as ``Test'', as indicated in Table \ref{estações_avaliadas}, are presented. We observe that these intervals maintain a consistent pattern, covering a wide range that encompasses all observed values. It is noteworthy that the amplitude of the intervals for the ``Test'' meteorological station, which was not used in model training and is located outside the geographic distribution of the observed data, is comparable to that of the ``Training with prediction'' station. Additionally, other analyses with different observations were conducted, and the results obtained were similar to those presented in this section.

\FloatBarrier
\section{Conclusion}\label{conclusao}

In this study, we introduced space-time regression models for functional data, incorporating clustering and repeated measures structure by utilizing spatial and temporal random components. These components were constructed through expansions using cubic B-spline basis functions, with random coefficients following normal distributions. The complexity of the model, with a large number of parameters, rendered maximum likelihood estimation inefficient. However, Bayesian parameter estimation in the model produced satisfactory results in simulation studies.

We evaluated the estimates of the parameter $\kappa$, which defines the order of the modified Bessel function of the third kind in the {\itshape{Matérn}} correlation structure, and the number of bases used in the functional structures representing fixed, spatial, and temporal effects, using model comparison criteria LPML and DIC7. These criteria effectively estimated the true value of the parameter $\kappa$. Preliminary simulation studies indicated that applying the same number of B-spline bases to all functional components of the model resulted in confounding issues in parameter estimation.

In the sensitivity study of prior distributions, we observed that the variance components of white noise remained stable regardless of the choice of priors, with consistent posterior estimates and coverage probabilities. However, the variance components of the spatial and temporal random effects showed less sensitivity when using the Inverse Gamma prior distribution with a shape parameter of $0.01$ and a scale parameter of $0.01$.

Following the selection of prior distributions, we conducted a simulation study which revealed that, within the evaluated perspective, increasing the data size, represented by the number of observed stations ($n$), analyzed months ($\tau$), and repeated measures ($J_i$, $i=1,2,3$), improved the accuracy in estimating the variance components of white noise. However, this improvement was not observed in the variance components of spatial and temporal effects. The algebraic structures of these components, which involve tensor products, made the parameter estimation process more complex, although this did not directly impact forecasting.

We applied the proposed model to precipitation data from 87 meteorological stations in the state of Goiás, Brazil, over a 21-year period (1980-2001), classified based on climatic conditions related to sea surface temperature in the Equatorial Pacific Ocean. The model selection techniques LPML and DIC7 led us to choose a configuration that encompasses the selection of $\kappa$ and the number of bases in the functional structures.

Analyzing the credibility intervals of the precipitation data, we found that the selected model generated credibility bands that covered most of the observed points. On the other hand, the prediction intervals showed high variability but also encompassed all observed points.

Therefore, the proposed model demonstrated promising results in simulations and in estimating precipitation in Goiás. Its functional flexibility allows for effective adaptation to data, enabling the modeling of nonlinear relationships. Describing spatial and temporal components through B-spline expansions facilitates capturing spatial and temporal correlations among observations within the same block, assigning the task of explaining this space-time correlation between repeated observations in a given block to the model's random error.

In summary, the results achieved in this study enable the creation of a precipitation grid, estimating the rainfall index in areas without available rain data, thereby supporting agricultural planning in regions of interest. The benefits include optimizing sowing dates based on accumulated precipitation, applying mechanistic models that simulate crop development, growth, and productivity, indicating new crops that can adapt to the region, and aiding in the watershed water management planning.

\bibliography{wileyNJD-APA}%

\begin{thebibliography}{}

\bibitem [\protect \citeauthoryear {%
Abramowitz%
\ \BBA {} Stegun%
}{%
Abramowitz%
\ \BBA {} Stegun%
}{%
{\protect \APACyear {1970}}%
}]{%
abramowitz1970handbook}
\APACinsertmetastar {%
abramowitz1970handbook}%
\begin{APACrefauthors}%
Abramowitz, M.%
\BCBT {}\ \BBA {} Stegun, I.%
\end{APACrefauthors}%
\unskip\
\newblock
\APACrefYearMonthDay{1970}{}{}.
\newblock
{\BBOQ}\APACrefatitle {Handbook of Mathematical Functions pover} {Handbook of mathematical functions pover}.{\BBCQ}
\newblock
\APACjournalVolNumPages{New York}{}{}{}.
\PrintBackRefs{\CurrentBib}

\bibitem [\protect \citeauthoryear {%
Aryaputera%
, Yang%
, Zhao%
\BCBL {}\ \BBA {} Walsh%
}{%
Aryaputera%
\ \protect \BOthers {.}}{%
{\protect \APACyear {2015}}%
}]{%
aryaputera2015very}
\APACinsertmetastar {%
aryaputera2015very}%
\begin{APACrefauthors}%
Aryaputera, A\BPBI W.%
, Yang, D.%
, Zhao, L.%
\BCBL {}\ \BBA {} Walsh, W\BPBI M.%
\end{APACrefauthors}%
\unskip\
\newblock
\APACrefYearMonthDay{2015}{}{}.
\newblock
{\BBOQ}\APACrefatitle {Very short-term irradiance forecasting at unobserved locations using spatio-temporal kriging} {Very short-term irradiance forecasting at unobserved locations using spatio-temporal kriging}.{\BBCQ}
\newblock
\APACjournalVolNumPages{Solar Energy}{122}{}{1266--1278}.
\PrintBackRefs{\CurrentBib}

\bibitem [\protect \citeauthoryear {%
Blangiardo%
, Cameletti%
, Baio%
\BCBL {}\ \BBA {} Rue%
}{%
Blangiardo%
\ \protect \BOthers {.}}{%
{\protect \APACyear {2013}}%
}]{%
blangiardo2013spatial}
\APACinsertmetastar {%
blangiardo2013spatial}%
\begin{APACrefauthors}%
Blangiardo, M.%
, Cameletti, M.%
, Baio, G.%
\BCBL {}\ \BBA {} Rue, H.%
\end{APACrefauthors}%
\unskip\
\newblock
\APACrefYearMonthDay{2013}{}{}.
\newblock
{\BBOQ}\APACrefatitle {Spatial and spatio-temporal models with {R}-INLA} {Spatial and spatio-temporal models with {R}-inla}.{\BBCQ}
\newblock
\APACjournalVolNumPages{Spatial and spatio-temporal epidemiology}{4}{}{33--49}.
\PrintBackRefs{\CurrentBib}

\bibitem [\protect \citeauthoryear {%
Celeux%
, Forbes%
, Robert%
\BCBL {}\ \BBA {} Titterington%
}{%
Celeux%
\ \protect \BOthers {.}}{%
{\protect \APACyear {2006}}%
}]{%
Celeux2006}
\APACinsertmetastar {%
Celeux2006}%
\begin{APACrefauthors}%
Celeux, G.%
, Forbes, F.%
, Robert, C.%
\BCBL {}\ \BBA {} Titterington, D.%
\end{APACrefauthors}%
\unskip\
\newblock
\APACrefYearMonthDay{2006}{}{}.
\newblock
{\BBOQ}\APACrefatitle {Deviance {Information} {Criteria} for {Missing} {Data} {Models}} {Deviance {Information} {Criteria} for {Missing} {Data} {Models}}.{\BBCQ}
\newblock
\APACjournalVolNumPages{Bayesian Analysis}{1}{4}{309–321}.
\PrintBackRefs{\CurrentBib}

\bibitem [\protect \citeauthoryear {%
Costa-Neto%
, Matta%
, Fernandes%
, Stone%
\BCBL {}\ \BBA {} Heinemann%
}{%
Costa-Neto%
\ \protect \BOthers {.}}{%
{\protect \APACyear {2023}}%
}]{%
costa2023environmental}
\APACinsertmetastar {%
costa2023environmental}%
\begin{APACrefauthors}%
Costa-Neto, G.%
, Matta, D\BPBI H\BPBI d.%
, Fernandes, I\BPBI K.%
, Stone, L\BPBI F.%
\BCBL {}\ \BBA {} Heinemann, A\BPBI B.%
\end{APACrefauthors}%
\unskip\
\newblock
\APACrefYearMonthDay{2023}{}{}.
\newblock
{\BBOQ}\APACrefatitle {Environmental clusters defining breeding zones for tropical irrigated rice in Brazil} {Environmental clusters defining breeding zones for tropical irrigated rice in brazil}.{\BBCQ}
\newblock
\APACjournalVolNumPages{Agronomy Journal}{}{}{}.
\PrintBackRefs{\CurrentBib}

\bibitem [\protect \citeauthoryear {%
Cressie%
\ \BBA {} Huang%
}{%
Cressie%
\ \BBA {} Huang%
}{%
{\protect \APACyear {1999}}%
}]{%
Matern1999}
\APACinsertmetastar {%
Matern1999}%
\begin{APACrefauthors}%
Cressie, N.%
\BCBT {}\ \BBA {} Huang, C.%
\end{APACrefauthors}%
\unskip\
\newblock
\APACrefYearMonthDay{1999}{}{}.
\newblock
{\BBOQ}\APACrefatitle {Classes of {Nonseparable}, {Spatio}-{Temporal} {Stationary} {Covariance} {Functions}} {Classes of {Nonseparable}, {Spatio}-{Temporal} {Stationary} {Covariance} {Functions}}.{\BBCQ}
\newblock
\APACjournalVolNumPages{Journal of the American Statistical Association}{94}{448}{631-647}.
\PrintBackRefs{\CurrentBib}

\bibitem [\protect \citeauthoryear {%
Cressie%
\ \BBA {} Wikle%
}{%
Cressie%
\ \BBA {} Wikle%
}{%
{\protect \APACyear {2015}}%
}]{%
cressie2015statistics}
\APACinsertmetastar {%
cressie2015statistics}%
\begin{APACrefauthors}%
Cressie, N.%
\BCBT {}\ \BBA {} Wikle, C\BPBI K.%
\end{APACrefauthors}%
\unskip\
\newblock
\APACrefYear{2015}.
\newblock
\APACrefbtitle {Statistics for spatio-temporal data} {Statistics for spatio-temporal data}.
\newblock
\APACaddressPublisher{}{John Wiley \& Sons}.
\PrintBackRefs{\CurrentBib}

\bibitem [\protect \citeauthoryear {%
Daly%
\ \protect \BOthers {.}}{%
Daly%
\ \protect \BOthers {.}}{%
{\protect \APACyear {2008}}%
}]{%
daly2008physiographically}
\APACinsertmetastar {%
daly2008physiographically}%
\begin{APACrefauthors}%
Daly, C.%
, Halbleib, M.%
, Smith, J\BPBI I.%
, Gibson, W\BPBI P.%
, Doggett, M\BPBI K.%
, Taylor, G\BPBI H.%
\BDBL {}Pasteris, P\BPBI P.%
\end{APACrefauthors}%
\unskip\
\newblock
\APACrefYearMonthDay{2008}{}{}.
\newblock
{\BBOQ}\APACrefatitle {Physiographically sensitive mapping of climatological temperature and precipitation across the conterminous {United} {States}} {Physiographically sensitive mapping of climatological temperature and precipitation across the conterminous {United} {States}}.{\BBCQ}
\newblock
\APACjournalVolNumPages{International Journal of Climatology: a Journal of the Royal Meteorological Society}{28}{15}{2031--2064}.
\PrintBackRefs{\CurrentBib}

\bibitem [\protect \citeauthoryear {%
De%
}{%
De%
}{%
{\protect \APACyear {1978}}%
}]{%
Boor1978}
\APACinsertmetastar {%
Boor1978}%
\begin{APACrefauthors}%
De, C., Boor.%
\end{APACrefauthors}%
\unskip\
\newblock
\APACrefYear{1978}.
\newblock
\APACrefbtitle {A {Practical} {Guide} to {Splines}} {A {Practical} {Guide} to {Splines}}.
\newblock
\APACaddressPublisher{New York}{Springer-Verlag}.
\PrintBackRefs{\CurrentBib}

\bibitem [\protect \citeauthoryear {%
Emetere%
}{%
Emetere%
}{%
{\protect \APACyear {2022}}%
}]{%
emetere2022numerical}
\APACinsertmetastar {%
emetere2022numerical}%
\begin{APACrefauthors}%
Emetere, M\BPBI E.%
\end{APACrefauthors}%
\unskip\
\newblock
\APACrefYear{2022}.
\newblock
\APACrefbtitle {Numerical Methods in Environmental Data Analysis} {Numerical methods in environmental data analysis}.
\newblock
\APACaddressPublisher{}{Elsevier}.
\PrintBackRefs{\CurrentBib}

\bibitem [\protect \citeauthoryear {%
Gamerman%
, Ippoliti%
\BCBL {}\ \BBA {} Valentini%
}{%
Gamerman%
\ \protect \BOthers {.}}{%
{\protect \APACyear {2022}}%
}]{%
gamerman2022dynamic}
\APACinsertmetastar {%
gamerman2022dynamic}%
\begin{APACrefauthors}%
Gamerman, D.%
, Ippoliti, L.%
\BCBL {}\ \BBA {} Valentini, P.%
\end{APACrefauthors}%
\unskip\
\newblock
\APACrefYearMonthDay{2022}{}{}.
\newblock
{\BBOQ}\APACrefatitle {A dynamic structural equation approach to estimate the short-term effects of air pollution on human health} {A dynamic structural equation approach to estimate the short-term effects of air pollution on human health}.{\BBCQ}
\newblock
\APACjournalVolNumPages{Journal of the Royal Statistical Society Series C: Applied Statistics}{71}{3}{739--769}.
\PrintBackRefs{\CurrentBib}

\bibitem [\protect \citeauthoryear {%
Gamerman%
\ \BBA {} Lopes%
}{%
Gamerman%
\ \BBA {} Lopes%
}{%
{\protect \APACyear {2006}}%
}]{%
Gamerman2006}
\APACinsertmetastar {%
Gamerman2006}%
\begin{APACrefauthors}%
Gamerman, D.%
\BCBT {}\ \BBA {} Lopes, H\BPBI F.%
\end{APACrefauthors}%
\unskip\
\newblock
\APACrefYear{2006}.
\newblock
\APACrefbtitle {Markov {Chain} {Monte} {Carlo} : {Stochastic} {Simulation} for {Bayesian} {Inference}} {Markov {Chain} {Monte} {Carlo} : {Stochastic} {Simulation} for {Bayesian} {Inference}}.
\newblock
\APACaddressPublisher{Boca Raton}{Chapman and Hall}.
\PrintBackRefs{\CurrentBib}

\bibitem [\protect \citeauthoryear {%
Geisser%
}{%
Geisser%
}{%
{\protect \APACyear {1993}}%
}]{%
Geisser1993}
\APACinsertmetastar {%
Geisser1993}%
\begin{APACrefauthors}%
Geisser, S.%
\end{APACrefauthors}%
\unskip\
\newblock
\APACrefYear{1993}.
\newblock
\APACrefbtitle {Predictive Inference} {Predictive inference}.
\newblock
\APACaddressPublisher{New York}{CRC Press}.
\PrintBackRefs{\CurrentBib}

\bibitem [\protect \citeauthoryear {%
Geisser%
\ \BBA {} Eddy%
}{%
Geisser%
\ \BBA {} Eddy%
}{%
{\protect \APACyear {1979}}%
}]{%
Geisser1979}
\APACinsertmetastar {%
Geisser1979}%
\begin{APACrefauthors}%
Geisser, S.%
\BCBT {}\ \BBA {} Eddy, W\BPBI F.%
\end{APACrefauthors}%
\unskip\
\newblock
\APACrefYearMonthDay{1979}{}{}.
\newblock
{\BBOQ}\APACrefatitle {A {Predictive} {Approach} to {Model} {Selection}} {A {Predictive} {Approach} to {Model} {Selection}}.{\BBCQ}
\newblock
\APACjournalVolNumPages{Journal of the American Statistical Association}{74}{365}{153–160}.
\PrintBackRefs{\CurrentBib}

\bibitem [\protect \citeauthoryear {%
Gelman%
\ \BBA {} Rubin%
}{%
Gelman%
\ \BBA {} Rubin%
}{%
{\protect \APACyear {1992}}%
}]{%
Gelman1992}
\APACinsertmetastar {%
Gelman1992}%
\begin{APACrefauthors}%
Gelman, A.%
\BCBT {}\ \BBA {} Rubin, D\BPBI B.%
\end{APACrefauthors}%
\unskip\
\newblock
\APACrefYearMonthDay{1992}{}{}.
\newblock
{\BBOQ}\APACrefatitle {Inference from iterative simulation using multiple sequences} {Inference from iterative simulation using multiple sequences}.{\BBCQ}
\newblock
\APACjournalVolNumPages{Statistical Science}{}{}{457--472}.
\PrintBackRefs{\CurrentBib}

\bibitem [\protect \citeauthoryear {%
Guan%
, Hsu%
, Wey%
\BCBL {}\ \BBA {} Tsao%
}{%
Guan%
\ \protect \BOthers {.}}{%
{\protect \APACyear {2009}}%
}]{%
guan2009modeling}
\APACinsertmetastar {%
guan2009modeling}%
\begin{APACrefauthors}%
Guan, B\BPBI T.%
, Hsu, H\BHBI W.%
, Wey, T\BHBI H.%
\BCBL {}\ \BBA {} Tsao, L\BHBI S.%
\end{APACrefauthors}%
\unskip\
\newblock
\APACrefYearMonthDay{2009}{}{}.
\newblock
{\BBOQ}\APACrefatitle {Modeling monthly mean temperatures for the mountain regions of Taiwan by generalized additive models} {Modeling monthly mean temperatures for the mountain regions of taiwan by generalized additive models}.{\BBCQ}
\newblock
\APACjournalVolNumPages{Agricultural and Forest Meteorology}{149}{2}{281--290}.
\PrintBackRefs{\CurrentBib}

\bibitem [\protect \citeauthoryear {%
Heinemann%
, da Matta%
, Fernandes%
, Fritsche-Neto%
\BCBL {}\ \BBA {} Costa-Neto%
}{%
Heinemann%
\ \protect \BOthers {.}}{%
{\protect \APACyear {2022}}%
}]{%
heinemann2022enviromic}
\APACinsertmetastar {%
heinemann2022enviromic}%
\begin{APACrefauthors}%
Heinemann, A\BPBI B.%
, da Matta, D\BPBI H.%
, Fernandes, I\BPBI K.%
, Fritsche-Neto, R.%
\BCBL {}\ \BBA {} Costa-Neto, G\BPBI M.%
\end{APACrefauthors}%
\unskip\
\newblock
\APACrefYearMonthDay{2022}{}{}.
\newblock
{\BBOQ}\APACrefatitle {Enviromic prediction is useful to define the limits of climate adaptation: A case study of common beans in {Brazil}} {Enviromic prediction is useful to define the limits of climate adaptation: A case study of common beans in {Brazil}}.{\BBCQ}
\newblock
\APACjournalVolNumPages{bioRxiv}{}{}{}.
\PrintBackRefs{\CurrentBib}

\bibitem [\protect \citeauthoryear {%
Heinemann%
\ \protect \BOthers {.}}{%
Heinemann%
\ \protect \BOthers {.}}{%
{\protect \APACyear {2024}}%
}]{%
heinemann2024climate}
\APACinsertmetastar {%
heinemann2024climate}%
\begin{APACrefauthors}%
Heinemann, A\BPBI B.%
, Stone, L\BPBI F.%
, Silva, G\BPBI C\BPBI C.%
, Matta, D\BPBI H\BPBI d.%
, Justino, L\BPBI F.%
\BCBL {}\ \BBA {} Silva, S\BPBI C\BPBI d.%
\end{APACrefauthors}%
\unskip\
\newblock
\APACrefYearMonthDay{2024}{}{}.
\newblock
{\BBOQ}\APACrefatitle {Climate drivers afecting upland rice yield in the central region of Brazil} {Climate drivers afecting upland rice yield in the central region of brazil}.{\BBCQ}
\newblock
\APACjournalVolNumPages{Pesquisa Agropecu{\'a}ria Tropical}{54}{}{e77222}.
\PrintBackRefs{\CurrentBib}

\bibitem [\protect \citeauthoryear {%
Laurini%
}{%
Laurini%
}{%
{\protect \APACyear {2019}}%
}]{%
laurini2019spatio}
\APACinsertmetastar {%
laurini2019spatio}%
\begin{APACrefauthors}%
Laurini, M\BPBI P.%
\end{APACrefauthors}%
\unskip\
\newblock
\APACrefYearMonthDay{2019}{}{}.
\newblock
{\BBOQ}\APACrefatitle {A spatio-temporal approach to estimate patterns of climate change} {A spatio-temporal approach to estimate patterns of climate change}.{\BBCQ}
\newblock
\APACjournalVolNumPages{Environmetrics}{30}{1}{e2542}.
\PrintBackRefs{\CurrentBib}

\bibitem [\protect \citeauthoryear {%
Lewis%
, Allen%
, Richardson%
\BCBL {}\ \BBA {} Holt%
}{%
Lewis%
\ \protect \BOthers {.}}{%
{\protect \APACyear {2006}}%
}]{%
lewis2006error}
\APACinsertmetastar {%
lewis2006error}%
\begin{APACrefauthors}%
Lewis, K.%
, Allen, J.%
, Richardson, A.%
\BCBL {}\ \BBA {} Holt, J.%
\end{APACrefauthors}%
\unskip\
\newblock
\APACrefYearMonthDay{2006}{}{}.
\newblock
{\BBOQ}\APACrefatitle {Error quantification of a high resolution coupled hydrodynamic-ecosystem coastal-ocean model: Part3, validation with {Continuous} {Plankton} {Recorder} data} {Error quantification of a high resolution coupled hydrodynamic-ecosystem coastal-ocean model: Part3, validation with {Continuous} {Plankton} {Recorder} data}.{\BBCQ}
\newblock
\APACjournalVolNumPages{Journal of Marine Systems}{63}{3-4}{209--224}.
\PrintBackRefs{\CurrentBib}

\bibitem [\protect \citeauthoryear {%
Menezes%
, Casaroli%
, Heinemann%
, Moschetti%
\BCBL {}\ \BBA {} Battisti%
}{%
Menezes%
\ \protect \BOthers {.}}{%
{\protect \APACyear {2022}}%
}]{%
menezes2022impact}
\APACinsertmetastar {%
menezes2022impact}%
\begin{APACrefauthors}%
Menezes, C\BPBI T.%
, Casaroli, D.%
, Heinemann, A\BPBI B.%
, Moschetti, V\BPBI C.%
\BCBL {}\ \BBA {} Battisti, R.%
\end{APACrefauthors}%
\unskip\
\newblock
\APACrefYearMonthDay{2022}{}{}.
\newblock
{\BBOQ}\APACrefatitle {The impact of gridded weather database on soil water availability in rice crop modeling} {The impact of gridded weather database on soil water availability in rice crop modeling}.{\BBCQ}
\newblock
\APACjournalVolNumPages{Theoretical and Applied Climatology}{147}{3}{1401--1414}.
\PrintBackRefs{\CurrentBib}

\bibitem [\protect \citeauthoryear {%
NOAA%
}{%
NOAA%
}{%
{\protect \APACyear {2019}}%
}]{%
cliamte2019}
\APACinsertmetastar {%
cliamte2019}%
\begin{APACrefauthors}%
NOAA.%
\end{APACrefauthors}%
\unskip\
\newblock
\APACrefYearMonthDay{2019}{}{}.
\newblock
{\BBOQ}\APACrefatitle {Historical ENSO episodes (1950–present): Cold and warm episodes by season. National Weather Service, Climate Prediction Center} {Historical enso episodes (1950–present): Cold and warm episodes by season. national weather service, climate prediction center}.{\BBCQ}.
\newblock
\begin{APACrefURL} \url{http://www.cpc.ncep.noaa.gov/products/analysis_monitoring/ensostuff/ensoyears_ERSSTv3b.shtml} \end{APACrefURL}
\PrintBackRefs{\CurrentBib}

\bibitem [\protect \citeauthoryear {%
Plummer%
\ \protect \BOthers {.}}{%
Plummer%
\ \protect \BOthers {.}}{%
{\protect \APACyear {2019}}%
}]{%
CODA}
\APACinsertmetastar {%
CODA}%
\begin{APACrefauthors}%
Plummer, M.%
, Best, N.%
, Cowles, K.%
, Vines, K.%
, Sarkar, D.%
, Bates, D.%
\BDBL {}Magnusson, A.%
\end{APACrefauthors}%
\unskip\
\newblock
\APACrefYearMonthDay{2019}{}{}.
\newblock
{\BBOQ}\APACrefatitle {coda: Output Analysis and Diagnostics for MCMC} {coda: Output analysis and diagnostics for mcmc}{\BBCQ}\ [\bibcomputersoftwaremanual].
\newblock
\APACrefnote{R package version 0.19-3}
\PrintBackRefs{\CurrentBib}

\bibitem [\protect \citeauthoryear {%
Ramsay%
\ \BBA {} Silverman%
}{%
Ramsay%
\ \BBA {} Silverman%
}{%
{\protect \APACyear {1993}}%
}]{%
ReS2002}
\APACinsertmetastar {%
ReS2002}%
\begin{APACrefauthors}%
Ramsay, J\BPBI O.%
\BCBT {}\ \BBA {} Silverman, B\BPBI W.%
\end{APACrefauthors}%
\unskip\
\newblock
\APACrefYear{1993}.
\newblock
\APACrefbtitle {Applied Functional Data Analysis: Methods and Case Studies} {Applied functional data analysis: Methods and case studies}.
\newblock
\APACaddressPublisher{New York}{Springer}.
\PrintBackRefs{\CurrentBib}

\bibitem [\protect \citeauthoryear {%
Resende%
, Chenu%
, Rasmussen%
, Heinemann%
\BCBL {}\ \BBA {} Fritsche-Neto%
}{%
Resende%
\ \protect \BOthers {.}}{%
{\protect \APACyear {2022}}%
}]{%
resende2022enviromics}
\APACinsertmetastar {%
resende2022enviromics}%
\begin{APACrefauthors}%
Resende, R\BPBI T.%
, Chenu, K.%
, Rasmussen, S\BPBI K.%
, Heinemann, A\BPBI B.%
\BCBL {}\ \BBA {} Fritsche-Neto, R.%
\end{APACrefauthors}%
\unskip\
\newblock
\APACrefYearMonthDay{2022}{}{}.
\newblock
{\BBOQ}\APACrefatitle {enviromics in plant breeding} {enviromics in plant breeding}.{\BBCQ}
\newblock
\APACjournalVolNumPages{Frontiers in Plant Science}{13}{}{935380}.
\PrintBackRefs{\CurrentBib}

\bibitem [\protect \citeauthoryear {%
Robert%
, Casella%
\BCBL {}\ \BBA {} Casella%
}{%
Robert%
\ \protect \BOthers {.}}{%
{\protect \APACyear {1999}}%
}]{%
robert1999monte}
\APACinsertmetastar {%
robert1999monte}%
\begin{APACrefauthors}%
Robert, C\BPBI P.%
, Casella, G.%
\BCBL {}\ \BBA {} Casella, G.%
\end{APACrefauthors}%
\unskip\
\newblock
\APACrefYear{1999}.
\newblock
\APACrefbtitle {Monte {Carlo} statistical methods} {Monte {Carlo} statistical methods}\ (\BVOL~2).
\newblock
\APACaddressPublisher{}{Springer}.
\PrintBackRefs{\CurrentBib}

\bibitem [\protect \citeauthoryear {%
Stauffer%
, Mayr%
, Messner%
, Umlauf%
\BCBL {}\ \BBA {} Zeileis%
}{%
Stauffer%
\ \protect \BOthers {.}}{%
{\protect \APACyear {2017}}%
}]{%
stauffer2017spatio}
\APACinsertmetastar {%
stauffer2017spatio}%
\begin{APACrefauthors}%
Stauffer, R.%
, Mayr, G\BPBI J.%
, Messner, J\BPBI W.%
, Umlauf, N.%
\BCBL {}\ \BBA {} Zeileis, A.%
\end{APACrefauthors}%
\unskip\
\newblock
\APACrefYearMonthDay{2017}{}{}.
\newblock
{\BBOQ}\APACrefatitle {Spatio-temporal precipitation climatology over complex terrain using a censored additive regression model} {Spatio-temporal precipitation climatology over complex terrain using a censored additive regression model}.{\BBCQ}
\newblock
\APACjournalVolNumPages{International Journal of Climatology}{37}{7}{3264--3275}.
\PrintBackRefs{\CurrentBib}

\bibitem [\protect \citeauthoryear {%
Stein%
}{%
Stein%
}{%
{\protect \APACyear {1999}}%
}]{%
stein1999interpolation}
\APACinsertmetastar {%
stein1999interpolation}%
\begin{APACrefauthors}%
Stein, M\BPBI L.%
\end{APACrefauthors}%
\unskip\
\newblock
\APACrefYear{1999}.
\newblock
\APACrefbtitle {Interpolation of spatial data: some theory for kriging} {Interpolation of spatial data: some theory for kriging}.
\newblock
\APACaddressPublisher{}{Springer Science \& Business Media}.
\PrintBackRefs{\CurrentBib}

\bibitem [\protect \citeauthoryear {%
Team%
}{%
Team%
}{%
{\protect \APACyear {2013}}%
}]{%
R}
\APACinsertmetastar {%
R}%
\begin{APACrefauthors}%
Team, R\BPBI C.%
\end{APACrefauthors}%
\unskip\
\newblock
\APACrefYearMonthDay{2013}{}{}.
\newblock
{\BBOQ}\APACrefatitle {R: A Language and Environment for Statistical Computing} {R: A language and environment for statistical computing}{\BBCQ}\ [\bibcomputersoftwaremanual].
\newblock
\APACaddressPublisher{Vienna, Austria}{}.
\newblock
\begin{APACrefURL} \url{http://www.R-project.org/} \end{APACrefURL}
\PrintBackRefs{\CurrentBib}

\bibitem [\protect \citeauthoryear {%
Thornton%
, Running%
\BCBL {}\ \BBA {} White%
}{%
Thornton%
\ \protect \BOthers {.}}{%
{\protect \APACyear {1997}}%
}]{%
thornton1997generating}
\APACinsertmetastar {%
thornton1997generating}%
\begin{APACrefauthors}%
Thornton, P\BPBI E.%
, Running, S\BPBI W.%
\BCBL {}\ \BBA {} White, M\BPBI A.%
\end{APACrefauthors}%
\unskip\
\newblock
\APACrefYearMonthDay{1997}{}{}.
\newblock
{\BBOQ}\APACrefatitle {Generating surfaces of daily meteorological variables over large regions of complex terrain} {Generating surfaces of daily meteorological variables over large regions of complex terrain}.{\BBCQ}
\newblock
\APACjournalVolNumPages{Journal of hydrology}{190}{3-4}{214--251}.
\PrintBackRefs{\CurrentBib}

\bibitem [\protect \citeauthoryear {%
Xavier%
, King%
\BCBL {}\ \BBA {} Scanlon%
}{%
Xavier%
\ \protect \BOthers {.}}{%
{\protect \APACyear {2016}}%
}]{%
xavier2016daily}
\APACinsertmetastar {%
xavier2016daily}%
\begin{APACrefauthors}%
Xavier, A\BPBI C.%
, King, C\BPBI W.%
\BCBL {}\ \BBA {} Scanlon, B\BPBI R.%
\end{APACrefauthors}%
\unskip\
\newblock
\APACrefYearMonthDay{2016}{}{}.
\newblock
{\BBOQ}\APACrefatitle {Daily gridded meteorological variables in Brazil (1980--2013)} {Daily gridded meteorological variables in brazil (1980--2013)}.{\BBCQ}
\newblock
\APACjournalVolNumPages{International Journal of Climatology}{36}{6}{2644--2659}.
\PrintBackRefs{\CurrentBib}

\end{thebibliography}



\end{document}